\RequirePackage{fix-cm}
\documentclass[twocolumn]{svjour3}          
\smartqed  

\usepackage[utf8]{inputenc}
\usepackage{epsfig}
\usepackage{graphicx}
\usepackage{fancybox}
\usepackage{amssymb}
\usepackage{stmaryrd}
\usepackage{a4wide}
\usepackage{subfigure}
\usepackage[noend]{algorithmic}
\usepackage{algorithm}
\usepackage{boxedminipage}

\usepackage[table]{xcolor}
\usepackage{multirow}

\usepackage{amsmath}
\usepackage{balance}
\usepackage{marvosym}

\begin{document}
\sloppy

\title{Approximate Similarity Search for Online Multimedia Services on Distributed CPU--GPU Platforms\thanks{E. Valle and R. Torres thank FAPESP for the financial support to this work. Preprint --- submitted for peer review.}}

\titlerunning{Approximate Similarity Search for Online Multimedia Services on CPU--GPU Platforms}        

\author{George Teodoro         
	\and Eduardo Valle 
	\and Nathan Mariano
	\and Ricardo Torres
	\and Wagner Meira Jr
	\and Joel H. Saltz 
}


\institute{G. Teodoro(\Letter) and J.H. Saltz \at
	Center for Comprehensive Informatics, Emory University, GA, USA \\
	\email{\{gteodor,jhsaltz\}@emory.edu}           
	\and
	E. Valle \at
	Recod Lab / DCA / FEEC, State University of Campinas, SP,  Brazil\\
	\email{dovalle@dca.fee.unicamp.br}
	\and
	N. Mariano and W. Meira Jr \at
	Department of Computer Science, Universidade Federal de Minas Gerais, MG, Brazil\\
	\email{\{nathanr,meira\}@dcc.ufmg.br}
	\and
	R. Torres \at
	Recod Lab / DSI / IC, State University of Campinas, SP,  Brazil\\
	\email{rtorres@ic.unicamp.br}
}

\date{Received: date / Accepted: date}

\maketitle

\begin{abstract}
Similarity search in high-dimentional spaces is a pivotal operation found a
variety of database applications. Recently, there has been an increase interest
in similarity search for online content-based multimedia services. Those
services, however, introduce new challenges with respect to the very large
volumes of data that have to be indexed/searched, and the need to minimize
response times observed by the end-users. Additionally, those users dynamically interact
with the systems creating fluctuating query request rates, requiring the search algorithm to adapt in order
to better utilize the underline hardware to reduce response times. In order to address
these challenges, we introduce hypercurves, a flexible framework for answering
approximate k-nearest neighbor (kNN) queries for very large multimedia
databases, aiming at online content-based multimedia services. Hypercurves
executes on hybrid CPU--GPU environments, and is able to employ those devices
cooperatively to support massive query request  rates. In order to keep the
response times optimal as the request rates vary, it employs a novel
dynamic scheduler to partition the work between CPU and GPU.  Hypercurves was
throughly evaluated using a large database of multimedia descriptors. Its
cooperative CPU--GPU execution achieved performance improvements of up to
30$\times$ when compared to the single CPU-core version. The dynamic work
partition mechanism reduces the observed query response times in about 50\%
when compared to the best static CPU--GPU task partition configuration. In
addition, Hypercurves achieves $superlinear$ scalability in distributed
(multi-node) executions, while keeping a high guarantee of equivalence with its
sequential version --- thanks to the proof of probabilistic equivalence, which
supported its aggressive parallelization design.

\end{abstract}

\section{Introduction}
\label{sec:intro}

Similarity search is the process of finding among objects stored in a
reference database, those nearest to a query object. In multimedia processing, both the query and the database objects are represented 
by a feature vector in a high-dimensional space. Several choices are available to establish the notion of distance, Euclidean distance being the most common. That operation is of fundamental
importance for several applications in content-based multimedia retrieval
services, which include not only search engines for web
images~\cite{Penatti:2012:CSG:2109696.2110003} but also image recognition on
mobile devices~\cite{springerlink:10.1007/s11263-011-0506-3}, real-time song
identification~\cite{DBLP:conf/ismir/ChandrasekharSR11}, photo tagging in
social networks~\cite{4562956}, recognition of copyrighted
material~\cite{Valle:2008:FIV:1410140.1410175} and many others.  Nowadays,
those services instigate an exciting scientific frontier and impel a
multimillionaire consumer market.

Though those services may appear extremely diverse, they are all founded upon
the use of descriptors, which extract feature vectors from multimedia documents, thus giving them a perceptually
meaningful geometry. The descriptors allow us to bridge the so called
``semantic gap'': the disparity between the amorphous low-level coding of
multimedia, e.g., image pixels or audio samples, and the complex high-level
tasks, e.g. classification or document retrieval, we need to perform.  
Looking for similar documents becomes
equivalent to looking for similar vectors.  The actual query processing may be
complex, consisting of several phases, but similarity search will often be the
first step and, because of the (in-)famous ``curse of dimensionality'', one of
the most expensive. 

The success of current content-based multimedia retrieval services depends on
their ability to handle extremely large and increasing volumes of data, and
keep the response times observed by the end-user low. The databases needed to process even a tiny fractions of the images available in the Web are larger than the storage capacity of most commodity single-user machines.
The great majority of indexing methods for similarity search, however, were designed to
execute sequentially, and are not able to take advantage of the aggregate power
in distributed environments. Moreover, classical distributed algorithms tend to ignore the 
response-time of processing each individual query, and to concentrate in providing maximum throughput for batches
of queries. That strategy clashes with the online nature of content-based multimedia retrieval services, because just like on other search engines, the waiting time observed by individual user requests
is critical. Moreover, online interaction between user and services 
creates large variations in the query rates submitted to the system, requiring those systems to adapt continuously to better exploit the available hardware and lower the
response times whenever possible. 

In order to address these challenges, in this work, we propose Hypercurves, a
concurrent index built upon the sequential multidimensional index
Multicurves~\cite{Valle:2008:HDI:1458082.1458181,Valle:2010:10.1109:TCE.2010.5606242}
and the concurrent execution environment Anthill~\cite{anthillII,hpdc10george}.
Multicurves addresses the challenges of approximate similarity queries for
multimedia services, including an optimizing scheduler that adapts the
parallelization regimen online to minimize query response-times under
fluctuating request loads. Hypercurves near-linear speedups and super-linear
scaleup on distributed environments rest upon the fact that Multicurves design fits
extremely well into the filter--stream execution model implemented in Anthill.
Nevertheless, the transition from sequential to parallel indexing remains very
challenging, and depends crucially on the ability of accessing independently
 each partition of the data. We demonstrate that this can be done efficiently,
while keeping the algorithms equivalent with very high probability
(Section~\ref{sec:equivalence}).

Hypercurves was published in preliminary form
in~\cite{Teodoro:2011:APA:2063576.2063651}, where we evaluated its performance
in CPU-only multi-core distributed machines. Though it performed extremely well
as compared to the sequential version, its performance remained constrained by
the compute-intensive task of evaluating distances between the query feature
vector and hundreds of candidate vectors. In this paper, we address that
shortcoming by redesigning Hypercurves for execution on heterogeneous
environments, comprising both CPUs and GPUs (graphical processing units). GPUs
are massively parallel and power-efficient processors, which have found a niche
as accelerators for regular compute-intensive applications. The utilization of
GPUs with Hypercurves, however, is very challenging, since we are interested in
providing low response times under online workloads that vary throughout
execution. GPUs, on the other hand, are fundamentally throughput-oriented,
because they are built as a large collection of low-frequency computing cores,
which are able to process a very large number of simple operations in parallel.  

In Hypercurves, therefore, queries dispatched for execution with a GPU may
observe higher average response times, as they are using a collection of 
less powerful GPU computing cores as compared to a CPU core. Thus, a carefully
scheduling is employed to decide the best partition of queries between CPUs and
GPUs in order to minimize the average query response times during the
execution. This problem is specially critical in Hypercurves as the best
partition is affected by the load of requests submitted to the system 
throughout the execution.  

In this paper, we address these challenges, obtaining a dramatic improvement on
top of the former version of
Hypercurves~\cite{Teodoro:2011:APA:2063576.2063651}. The techniques we use to
schedule tasks for Hypercurves are also generalizable to other online
applications that benefit from GPU accelerations. The key contributions
include:

\begin{enumerate}

\item An improved Hypercurves, able to employ GPUs concurrently to
answer a massive number of requests in very large databases;

\item A careful design and implementation of optimizations for the
parallelization in hybrid CPU--GPU environments, including cooperation CPU--GPU
execution and asynchronous execution between these devices, as detailed in
Section~\ref{sec:heterogeneous}. This version achieved speedups of about
30$\times$ as compared to the CPU-only single core version of the application;

\item A dynamic scheduler algorithm for hybrid CPU--GPU environments, which
employs both devices cooperatively in order to minimize response times, and is
able to adapt the application automatically under fluctuating request loads to
optimize response times. When compare to the best static partition, the scheduler obtained average query response times
up to 48\% smaller. 

\end{enumerate}

All contributions are throughly evaluated in a comprehensive set of experiments.

The remainder of the text is organized as follows. In the next section, 
we discuss content-based multimedia
services, examining how the problem of similarity search is critical to their success.
Section~\ref{sec:foundations} summarizes the algorithmic foundations of this
work, by presenting the sequential index Multicurves, and the parallel framework
Anthill, which is used to build our parallel index Hypercurves. Hypercurves
parallelization strategy is detailed in Section~\ref{sec:hypercurves} along
with an analytical proof of the probabilistic equivalence between Multi- and
Hypercurves.  Section~\ref{sec:heterogeneous} introduces progressively
sophisticated execution plans for Hypercurves on heterogeneous CPU--GPU
environments. Section~\ref{sec:minQuery} discusses scheduling considering
heterogeneous CPU--GPU environments, under online time constrained applications
as Hypercurves.  In Section~\ref{sec:results} we present an experimental
evaluation of the proposed scheme in many stress scenarios, proceeding to the
conclusions in Section~\ref{sec:conclusions}.

\section{Related work}
\label{sec:relatedwork}

In textual data, low-level representation is strongly coupled with
semantic meaning because the correlation between textual words and high-level
concepts is strong. 

In multimedia, by contrast, the low-level coding (pixels, samples, frames) is
extremely distant from the high-level semantic concepts needed to answer the
user queries, precipitating the much debated ``semantic gap''. In order to
overcome that difficulty, it is necessary to embed the multimedia documents in
a space where distances represent perceptual dissimilarities: that is the task
of \textbf{descriptors}. The descriptors are an essential first step towards
bridging the gap between the amorphous low-level coding and the high-level
semantic concepts. 

Multimedia descriptors are very diverse, including a large choice of
representations for perceptual properties that may help to understand the
documents. Those properties include shape, color and texture for visual
documents; tone, pitch and timbre for audio documents; flow and rhythm of
movement for moving pictures; and many others. The descriptor gives these
perceptual properties a precise representation, by encoding them into a
\textbf{feature vector}. That induces a geometric organization where
perceptually similar documents are given vectors near in the space, while
perceptually distinct documents are given vectors further apart. To establish
those distances, often a simple metric is employed, like the Euclidean or the
Manhattan, but sometimes more complex metrics are chosen~\cite{Penatti:2012:CSG:2109696.2110003}.

Especially in what concerns images and videos, the last decade witnessed the
ascent of descriptors inspired by Computer Vision, especially the so-called
\emph{local
descriptors}~\cite{10.1109/TPAMI.2005.188,Tuytelaars:2008:LIF:1391081.1391082},
with the remarkable success of SIFT~\cite{Lowe:2004:DIF:993451.996342}. As their name suggests, local
descriptors represent the properties of small areas of the images or videos (in
opposition to the traditional \emph{global descriptors}, which attempt to
represent the entire document in a single feature vector). Their success was
followed by the idea of using compact representations based on their
quantization using codebooks, in the so-called ``bag of visual words'' model,
which became  one of the main tools in the
literature~\cite{10.1109/CVPR.2010.5539963}. 

Regardless of the specific choice, the retrieval of similar feature vectors becomes
a cornerstone operation to almost all systems.  That operation can be used
either directly (many early CBIR systems were little more than a similarity
search engine attached to a descriptor
space~\cite{Smeulders:2000:PAMI:895972}), either indirectly (similarity search
may be part of a kNN classifier, it can retrieve a preliminary set of
candidates to be refined by a more computationally intensive classifier, etc.).
In one way or another, it remains a critical component, if the system is to be
used in real-world, large-scale  databases~\cite{Liu:2007:PR:262}.


We can formalize the problem of search with multimedia descriptors in the
framework of feature-based processing of similarity queries of Böhm et al.~\cite{Bohm:2001:SHS:502807.502809}.
The multimedia description algorithm corresponds to the feature vector
extraction, which, formally is a function $F$ that maps a space of multimedia objects
${Obj}$ into $d$-dimensional real vectors:

\begin{equation}
F: {Obj} \rightarrow \mathbb{R}^d 
\end{equation}

Now, the dissimilarity between two objects ${obj}_i \in {Obj}$ can be determined by
establishing the distance (e.g., Euclidean) between their feature vectors:

\begin{equation}
\operatorname{\Delta}({obj}_1, {obj}_2) = \lVert F(obj_1), F(obj_2) \rVert
\end{equation}

Given that dissimilarity between objects, we can establish several types of
similarity queries~\cite{Bohm:2001:SHS:502807.502809} (range, nearest neighbor, $k$ nearest neighbors,
inverse $k$ nearest neighbors, etc.). In this work, we are especially interested
in $k$ nearest neighbors queries (kNN, for short). Given a database $B
\subseteq {Obj}$ and a query $q \in {Obj}$, the $k$ nearest neighbors to $q$ in $B$
are the indexed set of the $k$ objects in $B$ closest to $q$:

\begin{equation}
\begin{split}
&\operatorname{kNN}(B, q, k) = \bigl\{ b_1, \ldots, b_k \in B \; \bigl\vert \; \forall i \leq k \\
& \forall b \in B\backslash\{b_1 \ldots, b_i\}, 
 \operatorname{\Delta}(q,b_i) \leq \operatorname{\Delta}(q,b) \bigr\}
\label{eq:knndef}
\end{split}
\end{equation}

That defines the \emph{exact} version of kNN search. As we will see, for
large-scale multimedia services, that definition will have to be relaxed to
account for approximate answers, which allows for dramatic gains in speed.

\subsection{Prior Art}
\label{sec:priorart}

Efficient query processing for multidimensional data has been pursued for at
least four decades, with a myriad of applications that go far beyond multimedia
feature vector matching. Those include satisfying multi-criteria searches, and searches with
spatial and spatiotemporal
constraints~\cite{DuMouza:2009:LIS:1612760.1612764,Fagin:2001:OAA:375551.375567,Pang:2010:EPE:1825238.1825263,Yiu:2009:MTD:1553321.1553325}.

An exhaustive review would be overwhelming and beyond the scope of this
article. The most comprehensive reference to the subject is the textbook of
Samet~\cite{Samet:2005:FMM:1076819}. The book chapters of
Castelli~\cite{CastelliChapter} and Faloutsos~\cite{FaloutsosChapter} provide a
less daunting introduction, more focused on content-based based retrieval for
images. Another comprehensive, if somewhat old, reference is the survey of Böhm
et al.~\cite{Bohm:2001:SHS:502807.502809}, which also provides an excellent
introduction to the theme, with a good formalization of similarity queries, the
principles involved in their indexing and their cost models. The book edited by
Shakhnarovich et al. \cite{Shakhnarovich:2006:NML:1197919} focuses on computer
vision and machine applications. In what concerns metric methods, which are
able to process non-vector features, as long as they are embedded in a metric
space, the essential reference is the textbook of Zezula et
al.~\cite{Zezula:2010:SSM:1951721}. Although already decade-old, the survey of
Chávez et al.~\cite{Chavez:2001:SMS:502807.502808} is also an excellent, comprehensive
introduction to similarity search in metric spaces. 

Despite the huge assortment of methods available, those of practical interest
in the context of large-scale content-based multimedia services are
surprisingly few.  Because of the ``curse of dimensionality'' (explained
below), methods that insist on exact solutions are only adequate for
low-dimensional spaces, while multimedia feature vectors often have hundreds of
dimensions. Most methods assume that the implementation uses shared main
memory (with cheap uniform random access), which cannot be the case on the very
large databases we want to address. Other methods, such as the ones based on clustering, have
prohibitively high index building times (with a forced rebuilding if the index
changes too much), being adequate only for moderate-size static databases.

The focus of multimedia retrieval and classification on approximate techniques
is not just a result of the technical challenge of treating high
dimensionalities. Multimedia descriptors are always
intrinsically approximate, due to the fact the relationship between the visual properties they encode and the high-level semantic concepts remains limited. In addition, descriptors almost always employ quantization and averaging to various degrees, making them approximate also in a numerical and statistical sense.
Therefore, insisting on exact techniques makes no sense. What
is needed is a good trade-off between precision and speed. 

Approximation in kNN search may imply different compromises: sometimes
it means finding elements not too far from the exact answers, i.e., guaranteeing
that the distance to the elements returned will bet up to a factor from the
distance to the correct elements; sometimes it means a bounded probability of
missing the correct elements. Sometimes, the guarantee offered is more complex
than that, for example, a bounded probability of finding the correct answer,
provided it is sufficiently closer to the query than the closest incorrect
answer~\cite{DBLP:conf/stoc/IndykM98}.

Approximation on a bounded factor is formalized as following: given a database $B
\subseteq {Obj}$ and a query $q \in {Obj}$, the $(1+\epsilon)$ $k$ nearest neighbors to $q$ in $B$
are an indexed set of objects in $B$ whose distance to the true kNN is at most a $(1+\epsilon)$ factor higher:

\begin{equation}
\begin{split}
\operatorname{\epsilon-kNN}&(B, q, k) = \Bigl\{ b_1, \ldots, b_k \in B \; \Bigl\vert \\
& \forall i \leq k, \; \bigl[ \forall b \in B\backslash\{b_1,\ldots,b_i\}, \\
& \operatorname{\Delta}(q,b_i) \leq (1+\epsilon)\operatorname{\Delta}(q,b) \bigr] \Bigr\}
\label{eq:epsknndef}
\end{split}
\end{equation}

Some methods might only guarantee such results with a probability bounded by some constant.
More often than not, however, practical approximative methods offer no formal 
guarantees, but just good empirical performance.

If perfect accuracy can be excused, the efficiency requirements remain very
challenging: the method should perform well for high-dimensional data (hundreds
of dimensions) in very large databases (at least millions of records); it must
adapt well to secondary-memory storage, which in practice means that few random
accesses should be performed; it should be dynamic, i.e., allow data insertion
and deletion without performance degradation.

A common pattern found in methods useful for large-scale multimedia is a
strategy of projecting the data onto different subspaces and creating
\textbf{subindexes} for each of those subspaces. The subindexes can be queried more or
less independently, and the results aggregated to find the final answer.

MEDRANK is one of those methods, which projects the data into several random
straight lines.  The one-dimensional position in the line is used to index the
data~\cite{Fagin:2003:ESS:872757.872795}. The method has an interesting
theoretical analysis, establishing that under certain hypotheses, rank
aggregation on straight line projections offers some (lax) bounds on
approximation error.  The techniques employed by the algorithm were extremely
well succeeded in moderately-dimensional multi-criteria databases, for which it
is still feasible to search for exact solutions. In those cases, many of the
choices are provably optimal~\cite{Fagin:2001:OAA:375551.375567}. For
high-dimensional multimedia information, however, the technique fails, mainly
due to the lack of correlation between the distance in the straight lines and
the distance in the high-dimensional
space~\cite{Valle:2008:FIV:1410140.1410175}.

Locality-sensitive hashing (LSH) uses locality-aware hashing functions to index
the data. The method uses several of those ``hash tables'' at once, to improve
reliability~\cite{DBLP:conf/stoc/IndykM98}. LSH is backed by an interesting
theoretical background, which allows predicting the approximation bounds for
the index, for a given set of parameters. The well-succeeded family of pStable
locality sensitive hash functions~\cite{Datar:2004:LHS:997817.997857} has
allowed LSH to directly index Euclidean spaces, and its geometric fundament is
also strongly based on the idea of projection onto random straight lines. LSH
works extremely well when one wants to minimize the number of distances to be
evaluated, and can count on uniform cost random access to the data. However, in
situations where the cost of accessing the data dominates the cost of
computation, its efficiency is compromised. The parameterization of LSH tends
to favor the use of a large number of hash functions (and thus subindexes),
which also poses a challenge for scalability.

An interesting family of solutions employs the fractal space-filling curves.
Like MEDRANK, those methods reduce multidimensional indexing to one-dimensional
indexing, but using more sophisticated projections.  The method upon which we
build our work, Multicurves, is one of those methods and it is explained in
more detail on Section~\ref{sec:multicurves}. Since that family of methods is
particularly related to our work, we focus our review on them.

\subsection{Indexing with Space-filling Curves}
\label{sec:spacefilling-related}

\begin{figure}
\begin{center}
    \includegraphics[width=0.49\textwidth]{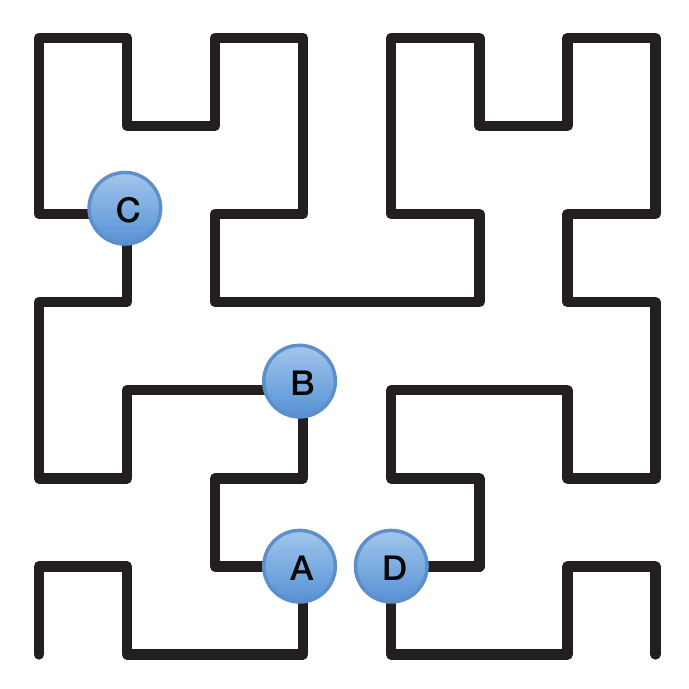}
\caption{Space-filling curves provide a ``vicinity-sensitive'' map: relative closeness in the space tends to be preserved in the curve (points A, B and C). However, in some boundary regions, those properties are violated (points A, B and D).}
\label{fig:spacefilling}
\end{center}
\end{figure}

Space-filling curves are maps from the unit interval to a hypercube of any
dimensionality~\cite{Sagan:1994}. Most of those curves are constructed by
fractal, self-similar, recursive procedures. Although the curves are
fascinating in themselves, here we are interested in their ability to induce a
``vicinity-sensitive'' total order in the data. With good probability, they
preserve neighborhood relations: if point A is closer to point B than to point
C in the space, that relationship tends to remain in the curve
(Figure~\ref{fig:spacefilling}).

Space-filling curves have been implicitly used to perform similarity searches
in multidimensional spaces for a very long time. Indeed, one of the first
multidimensional indexes ever proposed~\cite{MortonIndexing} employed them hidden in
the idea of ``bit shuffling'', ``bit interlacing'' or ``bit interleaving'',
which consisted in interleaving the bits of the individual space coordinates to
generate a search key. Interleaving the bits, in fact, induces a type of
space-filling curve called Z-order curve, which explains why the method works
well. However, it was Faloutsos~\cite{Faloutsos:1988:GCP:53064.53065} the first
to explicitly refer to the concept of curves, and were Faloutsos and
Roseman~\cite{Faloutsos:1989:FSK:73721.73746} the first to suggest the use of
curves other than the Z-order, first proposing the Gray-code curve and then the
Hilbert curve.

Those pioneering methods worked in a very simple way, using the curve to map the multidimensional vector onto a one-dimensional key representing the position in the curve (which we call here \emph{extended-key}). That position was then employed to perform the search by similarity. For example, when performing kNN search, a good heuristic is to take the nearest elements in the curve as the nearest elements in the space, because of the ``vicinity-sensitiveness'' explained above.

Unfortunately, points near in the space are not always near in the curve. In fact, the biggest problem when employing the curves is the existence of boundary regions where the neighborhood-relation preserving properties are violated, and points closer in space are placed further apart in the curve (Figure~\ref{fig:spacefilling}). That issue worsens dramatically as dimensionality grows~\cite{Liao:2001:HDS:645484.656398,Shepherd99afast}.

In order to conquer the boundary effects, Megiddo and Shaft~\cite{ms-ennib-97}
suggested the use of several curves at once. As is done for the multiple straight
lines of MEDRANK, or for the multiple hash-tables of LSH, we build an
independent subindex for each curve. The query is then sought on all subindexes, in
the hope that in at least one of them, it will not fall close to a boundary
region. Megiddo and Shaft present the idea in very general terms,
without describing which types of space-filling curves should be used and what
had to be done to make them different. Therefore, Shepherd et
al.~\cite{Shepherd99afast} developed that idea, specifically recommending the
use of several identical Hilbert curves, where different copies of the vectors
are transformed by random rotations and translations. Whether or not those
transformations could be optimized was left unanswered. Finally, Liao et
al.~\cite{Liao:2001:HDS:645484.656398} solved the problem of choosing the
transformations, by devising the necessary number of curves and an optimal set
of translations to obtain (lax) bounds on the approximation error in the case
of kNN search.

A depart from those methods was suggested by Mainar-Ruiz and Pérez-Cortés~\cite{Mainar-Ruiz:2006:ANN:1170748.1172124}. Instead of using
multiple curves, they propose using multiple instances of the same element in
only one curve. Before inserting those instances in the curve, the algorithm
disturbs randomly their position, to give them the opportunity of falling into different
regions of the curve. In that way even if the query falls in a problematic
region, chances are it will be reasonably near to at least one of the
instances. Akune et al. improved on that method, by proposing a more careful
placement of the instance copies on the curve~\cite{5597727}, obtaining thus a
significant improvement in precision.

Another depart was suggested by Valle et al. in
Multicurves~\cite{Valle:2008:HDI:1458082.1458181,Valle:2010:10.1109:TCE.2010.5606242},
which also employed several curves, but with the important difference that each
curve maps a projection of the vectors onto a moderate-dimensional subspace.
That dimensionality reduction makes for an efficient implementation, reducing
the effects of the ``curse of dimensionality''. Because of the exponential
nature of the “curse” it is more efficient to process several low or
moderate-dimensional indexes than a single high-dimensional one. That is
explained by the fact that we do not only gain the intrinsic advantages of
using multiple curves (i.e., elements that are incorrectly separated in one
curve will probably stay together in another), but also, we lower the boundary
effects inside each one of the curves. Multicurves is explained in detail in
Section \ref{sec:multicurves}.

\subsubsection{Technical Details}
\label{sec:spacefilling-details}

\begin{figure}
\begin{center}
    \includegraphics[width=0.49\textwidth]{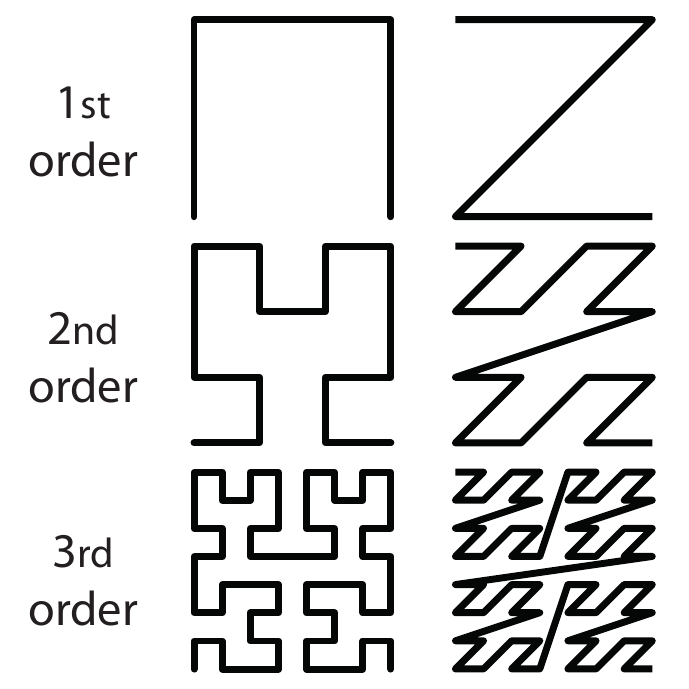}
\caption{Space-filling curves are usually obtained by a recursive refinement procedure, resulting in a fractal, self-similar curve. Each refinement step is called a \emph{order} of the curve. The curve fills the Continuum at the infinite limit of that procedure, but for working on digital data, it is not necessary to reach that limit. The figure shows three successive steps of the Hilbert curve (left) and the Z-order curve (right).}
\label{fig:spacefilling-orders}
\end{center}
\end{figure}

Space-filling curves are fractal curves introduced by G. Peano and D. Hilbert~\cite{Sagan:1994}, which provide a continuous surjective map $C:[0,1]\rightarrow[0,1]^{d}$ from the unit interval to a hypercube of any dimensionality. Most of those curves are constructed by  recursive procedures, where, in the limit, the curve fills the entire space (Figure~\ref{fig:spacefilling-orders}).

It was already known (due to a result of E. Netto) that such a mapping $C$ could not be at once bijective and continuous. Dropping injectivity, Peano was able to construct the first known continuous surjective map from the line to the space. Interestingly, in the recursive procedure to build the curve, all finite steps are bijective, but the limiting (and thus, effectively space-filling) curve becomes self-intersecting. In our applications with digital data, we can always consider $C$ bijective, because we never reach the limit needed to deal with the true Continuum.

When using the space-filling curve map in indexing, we are interested in the pre-images of the query and data points. Using the same notation as before, for $b_i \in B$ and $q \in {Obj}$, we are interested in those $C^{-1}(F(b_i))$ which are close to $C^{-1}(F(q))$ (remember that $F$ is the function that maps multimedia objects into feature vectors). We call the pre-image $C^{-1}(x)$ the \textbf{extended-key} of feature vector $x$.

There is a direct relationship between the number of refinements we need to go through in the recursive curve (called the \emph{order} of the curve) and the precision of the data we want to index. If we are employing dyadic curves (like the Z-order, Gray-order or Hilbert curves), we need $m^{th}$ order curves to index coordinates of $m$ bits. Remark that the bijective map $C^{-1}$ preserves the number of bits: from $d$ coordinates of $m$ bits each to a single extended-key with $d \times m$ bits.

It is important to emphasize that the curve does not have a concrete representation in the indexes. That is a common source of confusion for those who get acquainted for the first time with indexing based on space-filling curves. The curve is an useful abstraction, employed to create the map $C^{-1}$, which generates ``neighborhood-sensitive'' extended-keys. Then, the extended-keys are used in conventional, one-dimensional, indexing structures (a hash-table, a B-tree, etc.).

The actual computation of $C^{-1}$ depends, of course, on the type of space-filling curve being employed. For the Hilbert curve, several recursive algorithms have been proposed, but the most efficient scheme is an iterative one~\cite{ButzHilbert}. As we have mentioned, for the Z-order curve, the computation is extremely simple: it suffices to intercalate the bits of the coordinates.

It is interesting to analyze which kind of data can be indexed by the curves. The curves are able to organize vectors of any fixed-length ordinal data, provided that the order is the ``natural'' one: the order of the data is the same order of the numbers (binary codes) in which they are encoded. Otherwise, a transformation must be used to translate the vector of data into a vector of orders.

In the case of multimedia descriptors, we are mainly interested in vectors of numeric data. When the coordinates are integer, it is easy to see the scheme works, although the programmer must ensure to deal correctly with negative numbers in $C^{-1}$. Although less obvious to see, the scheme works with almost no modification for (IEEE 754) floating-point numbers. Indeed, because in that encoding the bits of the exponent are in more significant positions than the bits of the mantissa, the order of the encoded numbers is ``natural''. Again, the only caveat is to deal correctly with the most significant sign bit, used for negative numbers.

\section{Background}
\label{sec:foundations}

This section presents an introduction to Multicurves, the algorithmic foundation, upon which
our parallel solution is built; and details Anthill, the dataflow-based framework
employed in the parallelization.

\subsection{The Sequential Index Multicurves}
\label{sec:multicurves}

\begin{figure*}[htb!]
\begin{center}
    \includegraphics[width=\textwidth]{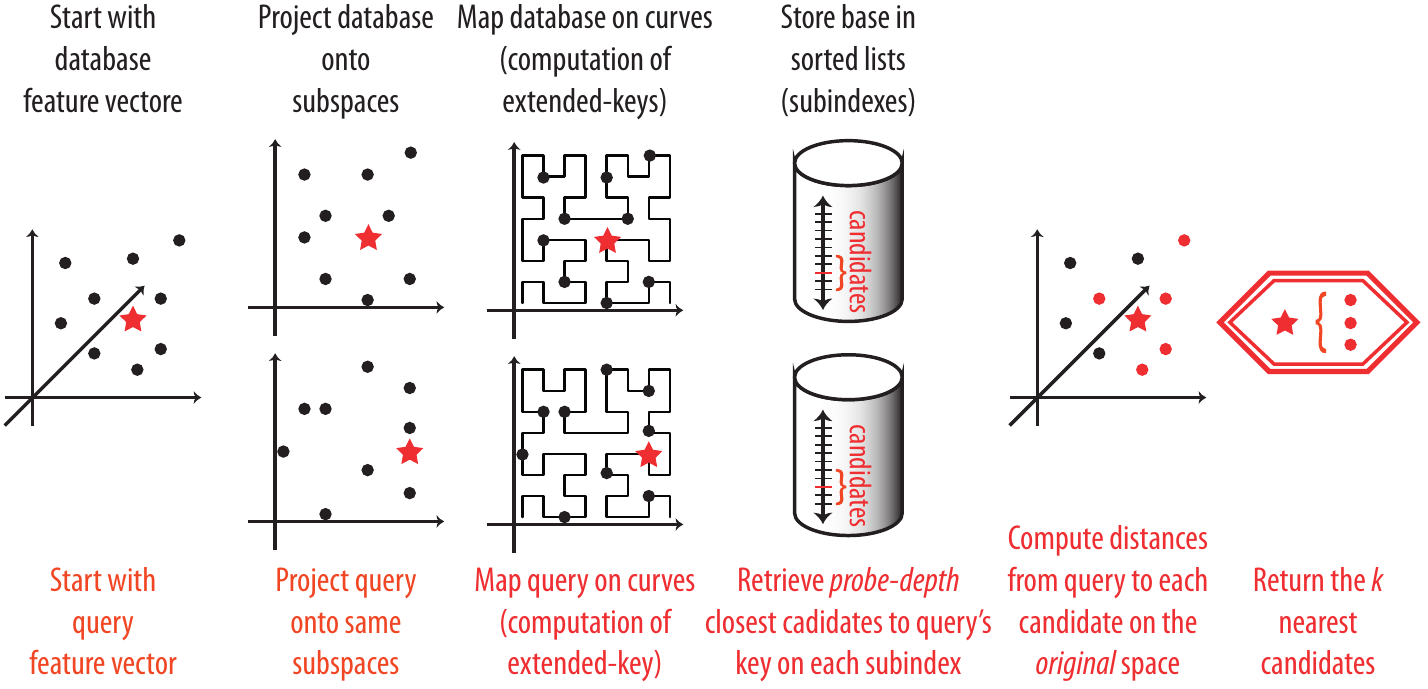}
\caption{Multicurves in action. (In black:)~The index is created by projecting the database feature vectors (small dots) onto different subspaces and mapping each projection in a space-filling curve to obtain the extended-keys. Each subspace induces an independent subindex, where the vectors are stored, sorted by extended-keys. (In red:)~Searching is performed by projecting the query feature vector (red star) onto the same subspaces and computing the extended-keys of the projections. A number (probe-depth) of candidates closest to the query's extended-key is retrieved from each subindex. Finally, the true distance of the candidates to the query is evaluated and the $k$ closest are returned.}
\label{fig:multicurves}
\end{center}
\end{figure*}

Multicurves~\cite{Valle:2008:HDI:1458082.1458181,Valle:2010:10.1109:TCE.2010.5606242}
is an index for accelerating kNN queries based on space-filling curves. Its properties make it
especially adapted for large-scale multimedia databases.

As we have seen in Section~\ref{sec:spacefilling-related}, the greatest problem
in using space-filling curves comes from boundary effects brought by the
existence of regions where their neighborhood-relation preserving properties are violated.
Different methods propose different solutions, usually through the simultaneous use of multiple
curves. As we have mentioned, Multicurves is also based on the use of multiple curves, but with the
important improvement that each curve is only responsible for a subset of the dimensions.
Because of the exponential nature of the ``curse'', it is more efficient to
process several low-dimensional queries than a single high-dimensional one. 

\begin{algorithm}
\begin{small}
\caption{Multicurves index construction}
\label{alg:buildMulticurves}
\flushleft
{\bf input:}
\begin{description}
	\item[${B}$:]       {the database elements to be indexed}
	\item[$\mathit{curves}$:]       {number of curves (and, thus, of subindexes)}
	\item[$\mathit{dims}{[}i{]}$:]  {dimensionality of the $i^{th}$ curve}
	\item[$a$:]         {attribution of the feature vectors dimensions to the subindexes --- $a[i,j]$ is the dimension in the $i^{th}$ subindex to which the $j^{th}$ dimension of the input data should be attributed (see line 7 below) }
	\item[$C^{-1}()$:]  {the space-filling curve map, as explained in Section~\ref{sec:spacefilling-details}}
	\item[$F()$:]       {the description function, returning a feature vector, as explained in Section~\ref{sec:relatedwork}. Usually, that will be already precomputed.}
\end{description}
{\bf output:} {an array of $\mathit{curves}$ sorted lists, which composes the index (each element is a subindex)}
\begin{algorithmic}[1]
	\STATE $\mathit{subindexes}[] \leftarrow$ {new array with} $\mathit{curves}$ {empty sorted lists};
	\FORALL{$b \in {B}$}
		\STATE ${v} \leftarrow F(b)$;
		\FOR{$c\leftarrow 1$ \textbf{to} $\mathit{curves}$}
			\STATE $\mathit{proj}[] \leftarrow$ {new array with} $\mathit{dims}[c]$ {empty elements};	
			\FOR{$d \leftarrow 1$ \textbf{to} $\mathit{dims}[c]$}
				\STATE $\mathit{proj}[d] \leftarrow v[a[c,d]]$;
			\ENDFOR
			\STATE $\mathit{key} \leftarrow C^{-1}(\mathit{proj})$;
			\STATE	{Insert} $<\mathit{key}, b>$ {into} $\mathit{subindexes}[\mathit{curve}]$;
		\ENDFOR
	\ENDFOR
	\RETURN $subindexes[]$;
\end{algorithmic}
\end{small}
\end{algorithm}

Multicurves index construction is simple
(Algorithm~\ref{alg:buildMulticurves}). The feature vector for each database
element is obtained (almost always, it will be computed beforehand, so the
operation in line 3 just retrieves the corresponding field). The dimensions of
the feature vectors are divided among a certain number of subindexes based on a
space-filling curve. Geometrically, that can be understood as projecting the
feature vector onto a subspace and then mapping it using a curve that fills the
subspace. For didactical reasons, the algorithm is presented as a ``batch''
operation, but nothing prevents the index from being built incrementally, as
long as the structure used to back the sorted lists allows so. 

\begin{algorithm}
\begin{small}
\caption{Multicurves search phase}
\label{alg:searchMulticurvesDist}
\flushleft
{\bf input :} (in addition to $\mathit{curves}$, $\mathit{dims}[]$, $a[]$, $C^{-1}()$ and $F()$ explained in Algorithm~\ref{alg:buildMulticurves})
\begin{description}
	\item[${k}$:]            {the number of desired nearest neighbors}
	\item[$\mathit{depth}$:] {the probe-depth, i.e., the number of elements to examine per subindex}
	\item[${q}$:]            {the data element to be queried}
	\item[$\mathit{subindexes}{[}{]}$:] {array of sorted lists composing the index, generated in Algorithm~\ref{alg:buildMulticurves})}	
\end{description}
{\bf output :} a list with the $k$ approximate nearest neighbors
\begin{algorithmic}[1]
		\STATE ${v} \leftarrow F(q)$;
   		\STATE $\mathit{candidates} \leftarrow \emptyset $;
		\FOR{$c\leftarrow 1$ \textbf{to} $\mathit{curves}$}
			\STATE $\mathit{proj}[] \leftarrow$ {new array with} $\mathit{dims}[c]$ {empty elements};	
			\FOR{$d \leftarrow 1$ \textbf{to} $\mathit{dims}[c]$}
				\STATE $\mathit{proj}[d] \leftarrow v[a[c,d]]$;
			\ENDFOR
			\STATE $\mathit{key} \leftarrow C^{-1}(\mathit{proj})$;
			\STATE $\mathit{candidates} \leftarrow \mathit{candidates} \; \cup \; \{ \mathit{depth}$ {closest vectors to} $\mathit{key}$ {in} $\mathit{subindex}[c] \}$;
		\ENDFOR
   		\STATE $\mathit{knn} \leftarrow \{ k$ {closest vectors to} $q$ {in} $\mathit{candidates} \}$;
	\RETURN $\mathit{knn}$ ;
\end{algorithmic}
\end{small}
\end{algorithm}

The search is conceptually similar: the query is decomposed into projections
(whose subspaces must be the same used during the index construction) and each
projection has its extended-key computed. Then, from each subindex, we obtain a
certain number of candidate elements ({\bf probe-depth}), whose extended-keys
are the nearest to the extended-key of the corresponding projection of the
query. In the end, we compute the actual distance from those elements to the
query and keep the $k$ nearest (Algorithm~\ref{alg:searchMulticurvesDist}).

The index creation and search processes are illustrated in Figure~\ref{fig:multicurves}.

It should be noted that in the scheme shown above, for simplicity sake, we supposed that both the query and the database elements are associated with a single feature vector by the description function $F()$. The extension of the algorithm is trivial for descriptors (like local descriptors) that generate several vectors per multimedia object, but bear in mind that the each vector is indexed and queried independently (for example, if a query object generates 10 feature vectors, the kNN search will produce 10 sets of $k$ nearest neighbors, one for each query vector). The task of taking a final decision (classification result, retrieval ranking) from those multiple answers is very application-dependent and beyond the scope of our article, which is concerned with the basic infra-structure. Here, we are concerned in achieving efficiently good results for each individual query vector.

In an experimental evaluation~\cite{Valle:2008:HDI:1458082.1458181,Valle:2010:10.1109:TCE.2010.5606242}
on high-dimensional feature vectors, Multicurves compared favorably to the state of the
art, represented by the methods of Liao et al.~\cite{Liao:2001:HDS:645484.656398} and Mainar-Ruiz and
Pérez-Cortes~\cite{Mainar-Ruiz:2006:ANN:1170748.1172124}, presenting a better
compromise between precision and speed.  It also performed well~\cite{Valle:2010:10.1109:TCE.2010.5606242}, when compared to
LSH~\cite{Datar:2004:LHS:997817.997857}, presenting an equivalent compromise
between precision and number of distances computed, but performing fewer random accesses.

\subsection{The Parallel Environment Anthill}
\label{sec:anthill}

Anthill~\cite{hpdc10george,anthillII,cluster09george,springerlink:10.1007/s10586-010-0151-6,anthill}
is a run-time system based on the filter--stream programming
model~\cite{datacutter} and, as such, applications are decomposed into
processing stages, called {\em filters}, which communicate with each other
using unidirectional {\em streams}.  At run time, Anthill spawns, on the nodes
of the cluster, instances of each filter, which are called transparent copies,
and automatically handles communication and state partitioning among those
copies~\cite{anthillII}.

\begin{figure}
\begin{center}
    \includegraphics[width=0.49\textwidth]{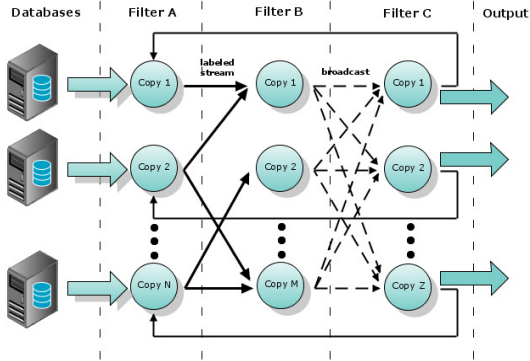}
\caption{The architecture of an Anthill application. Filters (columns) cooperate to process the data. Their communication is mediated by unidirectional streams (arrows). The filters are instantiated in transparent copies (circles) automatically by Anthill's runtime. The non-blocking I/O flow and event scheduling is also handled by Anthill. }
\label{fig:modelAndArch}
\end{center}
\end{figure}

When developing an application using the filter--stream model, both task and data 
parallelism are exploited. Task parallelism is achieved as the application is
broken up into a set of filters, which perform independently, accomplishing the application functionality
in a pipeline fashion. Data parallelism, on the other hand, is
obtained by creating transparently multiple copies of each filter and
distributing the data to be processed among them (Figure~\ref{fig:modelAndArch}).

Anthill provides an event-oriented filter programming abstraction, deriving
heavily from the message-oriented programming
model~\cite{coyote,xkernel,welsh01sosp}.  The streams that establish the
communication between filters generate input events, which must be handled.
The programmer provides handling functions for those events.  Anthill runtime
instantiates those functions and controls the non-blocking I/O flow to keep the
system running. It is a dataflow model, where event handling amounts to
asynchronous and independent tasks. Because the filters are multithreaded,
multiple tasks can be spawned when there are enough pending events and
computational resources. That feature is essential both in exploiting the full
capability of current multi-core architectures, and in spawning tasks on
multiple devices in heterogeneous, CPU--GPU equipped platforms. That flexibility
is accomplished by allowing the programmer to provide, for the same event,
handler functions targeting different devices which can be invoked by the
scheduler to use the appropriate processor.

\begin{figure}
\begin{center}
        \includegraphics[width=0.48\textwidth]{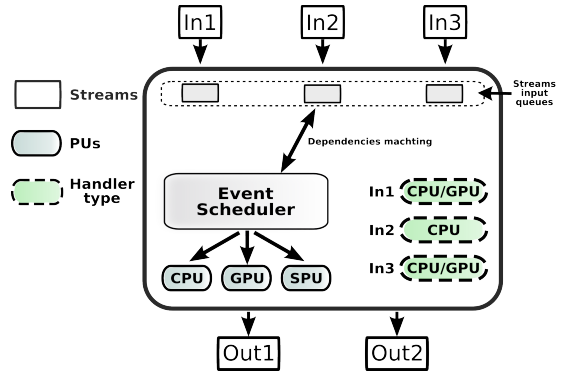}
\caption{The architecture of a single filter. Input streams (top blocks) generate events that must be handled by the filter. Different handler functions (dashed round boxes) can be provided by the programmer for each type of event and processing unit. The event scheduler coordinates the filter operation, dequeuing the input events and invoking the handling functions according to the available processing units (round boxes). As processing progresses, data is sent to the output streams (bottom blocks), generating events on the next filter (not shown). }
\label{fig:ahFilterArchitecure}
\end{center}
\end{figure}

Figure~\ref{fig:ahFilterArchitecure} illustrates the architecture of a typical
filter (a single application will be composed of several of those). It receives data from multiple input streams (\textit{In1, In2, and
In3}), each generating its own event queue, with handler functions
associated to each of them.  As shown, those functions are implemented
targeting different types of processors. The {\em Event
Scheduler}, depicted in the picture is responsible for consuming events from
the queues, invoking appropriate handlers according to the availability of
computational resources. As events are consumed, eventually some data is generated on
the filter that will be forwarded to the next filter.  As those data arrive in the next filter, they will trigger input events there. All filters run in parallel. Communication between filters, although not shown in the figure, is also managed by the run-time system.

When events are queued, they are not immediately assigned to a processor.  Rather, that occurs
on-demand, as devices become idle and request new events to process.  In the current
implementation, the demand-driven, first-come, first-served (DDFCFS) task
assignment policy is used as default strategy of the Event
Scheduler.

The first decision for the DDFCFS policy is to select from which queue to
execute events; this decision is made in a round-robin fashion, provided the
event has a handling function compatible with the available processor.  The
oldest event on the selected queue is dispatched for processing.  That simple
approach guarantees assignment to different devices according to their relative
performance in a transparent way, as processors will consume events in
proportion to their capacity to process them.

\section{The Distributed Index Hypercurves}
\label{sec:hypercurves}

In this section, we discuss how Multicurves (Section~\ref{sec:multicurves}) has been redesigned for
efficient execution on distributed environments, focusing on the CPU-only
version of the application (details of the GPU-based  presented in Section~\ref{sec:heterogeneous}).

Further,  we present a proof of the probabilistic equivalence between Multicurves
and Hypercurves (Section~\ref{sec:equivalence}). The proof is essential for the efficiency of the scheme, because in Hypercurves, 
the database is partitioned without overlapping among the nodes in the execution environment. 
Search is performed locally in the subindexes managed by each node, and a reduction stage merges the results. The cost of the algorithm
is dominated by the local searches, which are further dependent on the
probe-depth used (the number of candidates to retrieve from each subindex).
When using the same probe-depth of the sequential algorithm for each local
index of the distributed environment, the answer of Hypercurves is guaranteed
at least as good as the sequential algorithm. However, that is an extremely
pessimistic and costly choice for the local indexes probe-depth: we
show that the quality of Hypercurves is equivalent to that of Multicurves with
very high probability, when using a probe-depth slightly higher than the
original probe-depth divided by the number of nodes.

The ability to avoid data replication improves the scalability of the solution. The
user can further modify the probe-depth of the parallel algorithm according
to Equation 5 (Section~\ref{sec:equivalence}), to guarantee that the quality
of Hypercurves is equivalent to that of Multicurves with any desired probability.

\subsection{Hypercurves Parallelization Strategy}

Hypercurves~\cite{Teodoro:2011:APA:2063576.2063651} is a concurrent index built
upon the sequential multidimensional index
Multicurves~\cite{Valle:2010:10.1109:TCE.2010.5606242} and the concurrent
execution environment Anthill~\cite{anthillII}, in order to provide approximate
similarity search support for large-scale online multimedia services.
Therefore, Hypercurves addresses both the need to scale the database to
sizes beyond the capability of a single machine, and the need to keep the
answer times as short as possible.

Hypercurves strategy is to partition the database among the nodes (\emph{filter
copies} in the nomenclature of Anthill) of the distributed environment. The
queries are broadcast to all filter copies, which find a local answer in their
database subsets. The local answers are then reduced to a global answer in a
later merge step.  

\begin{figure*}[htb!]
\begin{center}
        \includegraphics[width=\textwidth]{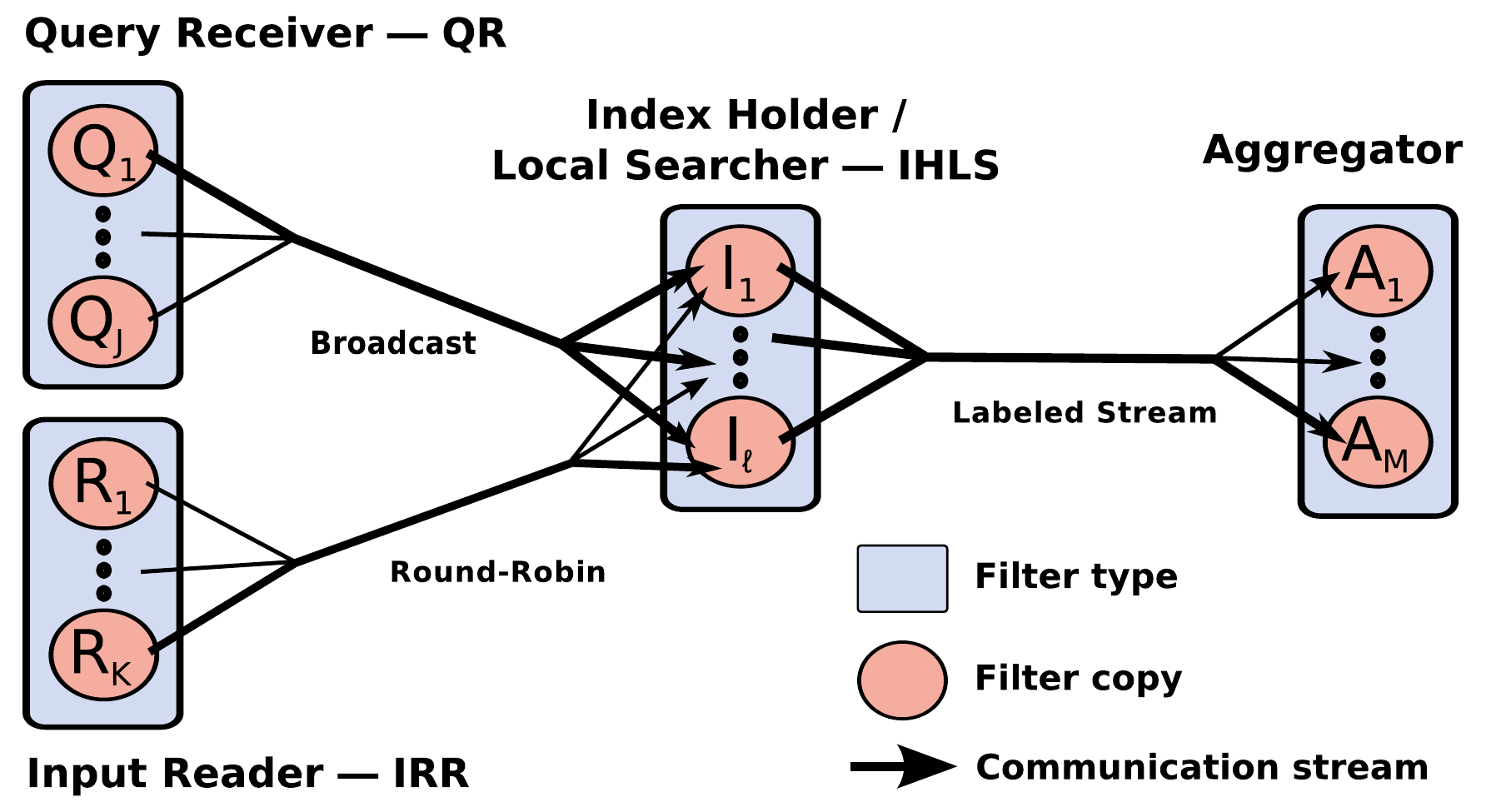}
\caption{Hypercurves parallelization design. Four filter types are involved: IRR, which reads data from the database and dispatches them to the IHLS to be indexed; QR, which reads queries from the user and dispatches them to the IHLS to be processed; IHLS, which provides a ``local'' index and query processing, for a subset of the data; Aggregator, which collects local kNN answers to the queries and aggregates them into a global kNN answer. Transparent copies of those filters are instantiated as needed by Anthill's runtime. Several types of streams are used in the communication between those copies: for example, during search, a query is broadcast from QR to all copies of IHLS; then all local answers relative to that query are sent to the same Aggregator filter, using the ``labeled stream'' facility.}
\label{fig:pmulticurves}
\end{center}
\end{figure*}

To better exploit Anthill execution environment, Hypercurves employs four types
of filter, organized in two parallel computation pipelines
(Figure~\ref{fig:pmulticurves}).  

The first pipeline is conceptually an index builder/updater, with the filters
\emph{Input Reader} (IRR) and \emph{Index Holder/Local Searcher} (IHLS).  IRR
reads the feature vectors from the input database and partitions them among the copies
of IHLS, which add the vectors received to their local index, according to
Algorithm~\ref{alg:buildMulticurves}. The filters execute concurrently, and
after the input is exhausted, interact to update the database.

The second pipeline, which is conceptually the query processor, contains three
filters: (i) \emph{Query Receiver} (QR); (ii) IHLS (shared with the first
pipeline); and (iii) \emph{Aggregator}. QR is the entry point to the search
server, receiving and broadcasting the queries to all IHLS copies.  For each
query, IHLS instances independently perform the search on their local index
partition, retrieving $k$ nearest \emph{local} feature vectors just like the
sequential Multicurves (Algorithm~\ref{alg:searchMulticurvesDist}). The final
answer is obtained by the Aggregator filter, which reduces the local answers
into $k$ \emph{global} nearest vectors. Since several Aggregator filter copies
may exist, it is crucial that the messages related to a particular query (same
query-id) be sent to the same Aggregator instance.  That is guaranteed by
making full use of Anthill Labeled-Stream communication policy, which computes
the particular copy of the Aggregator filters that will receive a given message
sent from IHLS based on a $hash$ computed in the query-id. Therefore, in this
context, query-id corresponds to the label of the message. The transaction
between IHLS and Aggregator is very similar to a generalized parallel data
reduction~\cite{Yu:2000:ARP:335231.335238}, except that it outputs a list of
values for each output, and that an arbitrary number of reductions are executed
in parallel. 

Hypercurves exploits all four dimensions of parallelism: task, data, pipeline,
and intra-filter. Task parallelism occurs as the two pipelines are executed in
parallel (e.g., index updates and searches).  Data parallelism is achieved as
the database is partitioned among the IHLS filters copies. Pipeline parallelism
results from Anthill ability to execute in parallel the filters of a single
computational pipeline (e.g., IRR and IHLS for updating the index).
Intra-filter parallelism refers to a single filter copy being able to process
events in parallel, thus, efficiently exploiting modern multi- and many-core computers. 


The broadcast from QR to IHLS has little impact on performance, because the
cost is dominated by the local searches. Therefore, the communication latency
is offset by the computation speedups. The disproportionate cost of those
searches has prompted a GPU-based implementation (Section~\ref{sec:heterogeneous}),
which in turn raised interesting challenges to the scheduling of the pipelined
events, leading to another important contribution of this work
(Sections~\ref{sec:minQuery}).

The cost of local searches depends critically on the probe-depth used (the
number of candidates to retrieve from each subindex). Hypercurves can be made
assuredly equivalent to Multicurves, by employing on each parallel node a
probe-depth at least as large as the one used in the sequential algorithm.
However, this over-pessimistic choice is unnecessarily costly and can be
significantly improved, as we will see in the next section.

\subsection{Probabilistic Equivalence Multicurves--Hypercurves}
\label{sec:equivalence}

Multicurves is based upon the ability of space-filling curves giving a total
order to data.  That makes each subindex a sorted list where a number of
candidates can be retrieved and then verified against the query in order to
obtain the $k$ nearest (Algorithm~\ref{alg:searchMulticurvesDist}).

In Hypercurves, the index is fragmented, with each IHLS filter copy having
available only a subset of the database: a single filter cannot warrant the
equivalent approximate $k$ nearest neighbors. That is the role of the
Aggregator filter: collecting the local best answers and returning a final
solution. 

In terms of equivalence between Multi- and Hypercurves it matters little how
the candidates are distributed among the IHLS instances, because the reduction
steps performed after the candidates are selected are conservative: they will
never discard one of the ``good'' answers once it is retrieved. Either Multi-
or Hypercurves will only miss a correct answer if they fail to retrieve it from
the subindexes. Therefore, Hypercurves can be made guaranteedly at least as
good as Multicurves by employing on each IHLS filter copy the same probe-depth
used on the sequential Multicurves. However, that is costly, and, as we will
shortly see, over-pessimistic.

\begin{figure}
\begin{center}
    \includegraphics[width=0.49\textwidth]{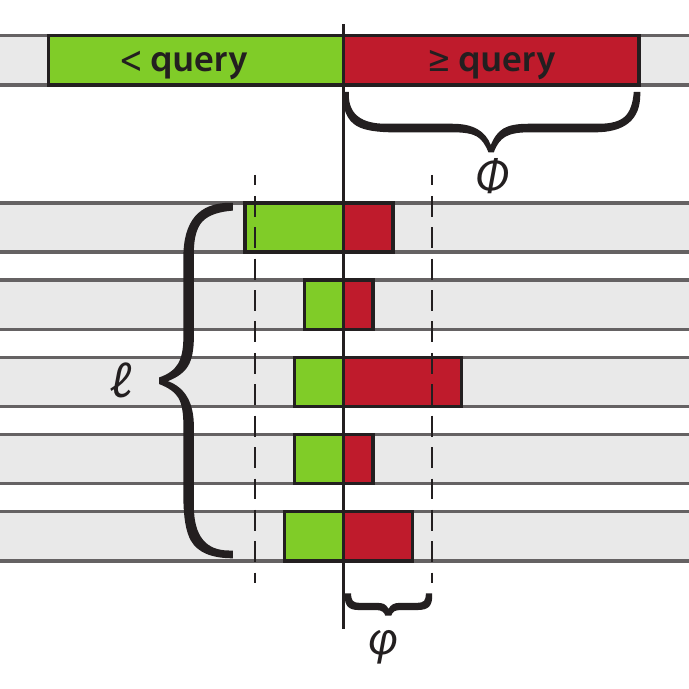}
\caption{The probabilistic equivalence between Multi- and Hypercurves corresponds to the following model. In a sorted list, for the query $q$, we retrieve $\Phi$ elements $<q$ and $\Phi$ elements $\geq q$. If we distribute the elements of that list randomly into $\ell$ sorted lists, how many $2\varphi$ elements must we retrieve in each of those new lists, in order to ensure missing none of the original ones. Because the elements  $<q$ and $\geq q$ cannot exchange positions, each ``half-list'' can be analyzed independently. In the example shown, the equivalence is not guaranteed, because some elements ``spill over'' the $\varphi$ limit in two of the half-lists.} 
\label{fig:proofmodel}
\end{center}
\end{figure}

Consider the same database, either in a Multicurves' subindex with
probe-depth~$= 2\Phi$, or partitioned among $\ell$ Hypercurves' IHLS filter
copies, each with probe-depth~$= 2\varphi$ (even probe-depths make the analyses
more symmetric, although the argument is essentially the same for odd values).
For any query, the candidates that would be in a single sorted list in
Multicurves are now distributed among $\ell$ sorted lists in Hypercurves.  In
more general terms, we start with a single sorted list and retrieve the $2\Phi$
elements closest to a query vector. If we distribute randomly that single sorted
list into $\ell$ sorted lists, how many elements must we retrieve from each of
those new lists (i.e. which value for $2\varphi$ must we employ) to ensure that
none of the originally retrieved elements is missed? Note that: (i)~due to the
sorted nature of the list, the elements before the query cannot exchange
positions with the elements after the query; (ii)~no element of the original
list can be lost as long as all those $2\ell$ ``half-lists'' are shorter than
$\varphi$. Those observations, which are essential to understand the
equivalence proof, are illustrated in Figure~\ref{fig:proofmodel}.

\begin{figure}[ht!]
\includegraphics[width=0.5\textwidth]{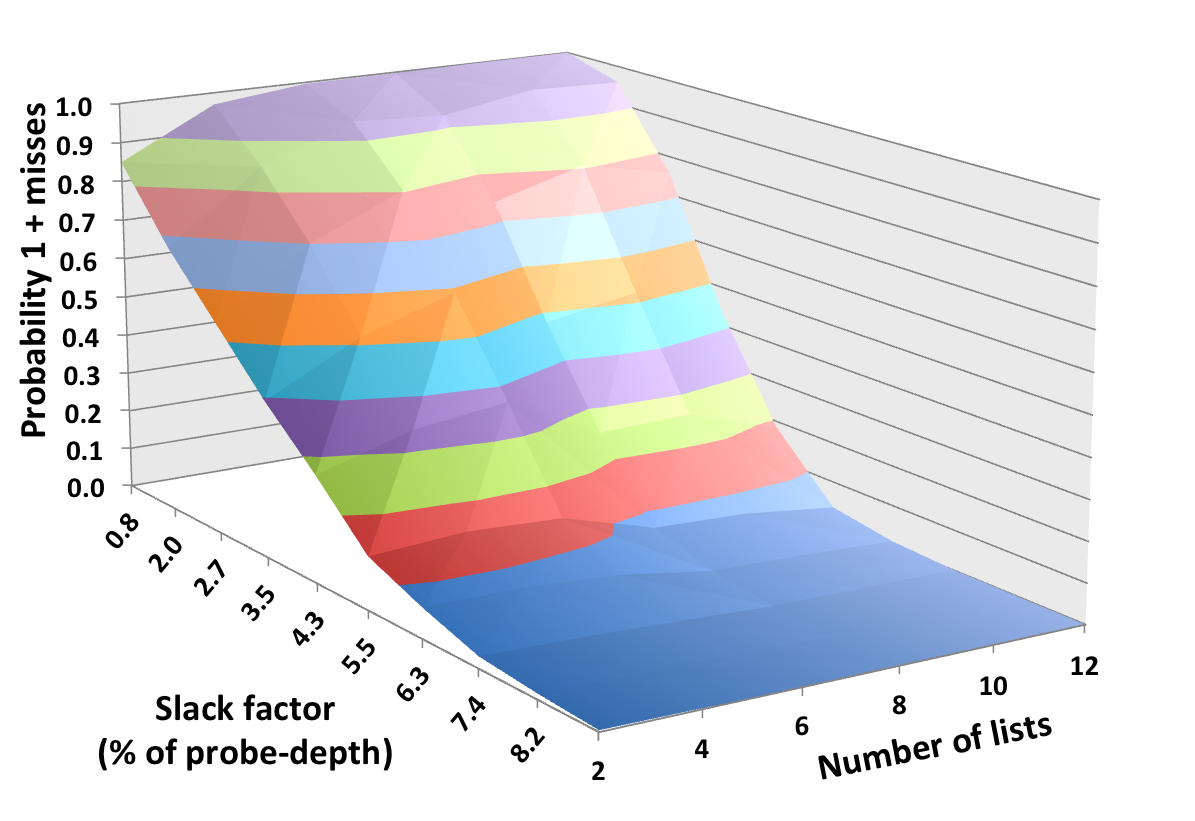}
\caption{Equivalence between sequential Multicurves with a probe-depth of $2\Phi=256$ and parallel Hypercurves with distributed probe-depth of $2\varphi$, with $\varphi=\left(1+\varsigma\right)\lceil\Phi/\ell\rceil$ and $\ell$~= the number of filter copies. The probability of missing any of the candidate vectors drops sharply to zero, for values of $\varsigma$ that are still very small.}
\label{fig:proof-equivalence}
\end{figure}

Due to (i), we can analyze each half of the list independently. The
distribution of the elements among the $\ell$ lists is given by a Multinomial
distribution with $\Phi$ trials and all probabilities equal to $\ell^{-1}$. The
exact probability of no list being longer than $\varphi$ involves computing a
truncated part of the distribution, but the exact formulas are exceedingly
complex and little elucidative. We can, however, bound it from
below~\cite{Mallows:1968:Biometrika:55} with:

\begin{equation}
P\left(List_{max}\leq\varphi\right)\geq 1-\left(\Phi\times P\left(List_{i}>\varphi\right)\right)
\end{equation}

\noindent where $List_{i}$ is an arbitrary single component of the equiprobable
Multinomial, which, by construction has a Binomial distribution for $\Phi$
trials and success rate of $\ell^{-1}$. Thus, the probability of any miss on
any of the $2\ell$ half-lists is bounded from above by:

\small
\begin{equation}
\begin{split}
&1-\operatorname{Max}\left[0; 1-\Phi\sum\limits_{k=\varphi+1}^{\Phi}\binom{\Phi}{k}\left(\frac{1}{\ell}\right)^k\left(1-\frac{1}{\ell}\right)^{\Phi-k}\right]^2 \\
& = 1-\operatorname{Max}\left[0; 1-\Phi\left(1-I_{1-\ell^{-1}}\left(\Phi-\varphi, \varphi+1\right)\right)\right]^2
\end{split}
\end{equation}
\normalsize

\noindent where $I()$ is the regularized incomplete Beta function. That
probability tends to zero for very reasonable values of $\varphi$, still much
lower than $\Phi$. That is more easily seen if we make
$\varphi=\left(1+\varsigma\right)\lceil\Phi/\ell\rceil$, i.e., if we
``distribute'' the probe-depth among the filters, adding a ``slack factor'' of
$\varsigma$. For all reasonable scenarios, the probability tends to zero very
fast, even for small $\varsigma$ (Figure~\ref{fig:proof-equivalence}). 

\section{Hypercurves in Heterogeneous CPU--GPU Environments}
\label{sec:heterogeneous}

The use of GPUs as general computing processors is a strong trend in high
performance computing, and represents a major paradigm shift towards massively
parallel and power efficient systems. GPus have an impressive computing power,
but taking advantage of them is challenging, especially for online services like Hypercurves. 

In this section, we introduce the design and implementation of Hypercurves for
heterogeneous, CPU--GPU environments, with a set of optimizations to maximize
its performance. We anticipate that the use of GPUs in this context raises
important challenges, especially in what concerns the optimization of response
times under fluctuating request loads. Those dynamic aspects are discussed in
Section~\ref{sec:minQuery}.

\subsection{GPU-based IHLS Implementation}
\label{sec:gpuIHLS}

\begin{figure*}[htp!]
\begin{center}
    \includegraphics[width=1\textwidth]{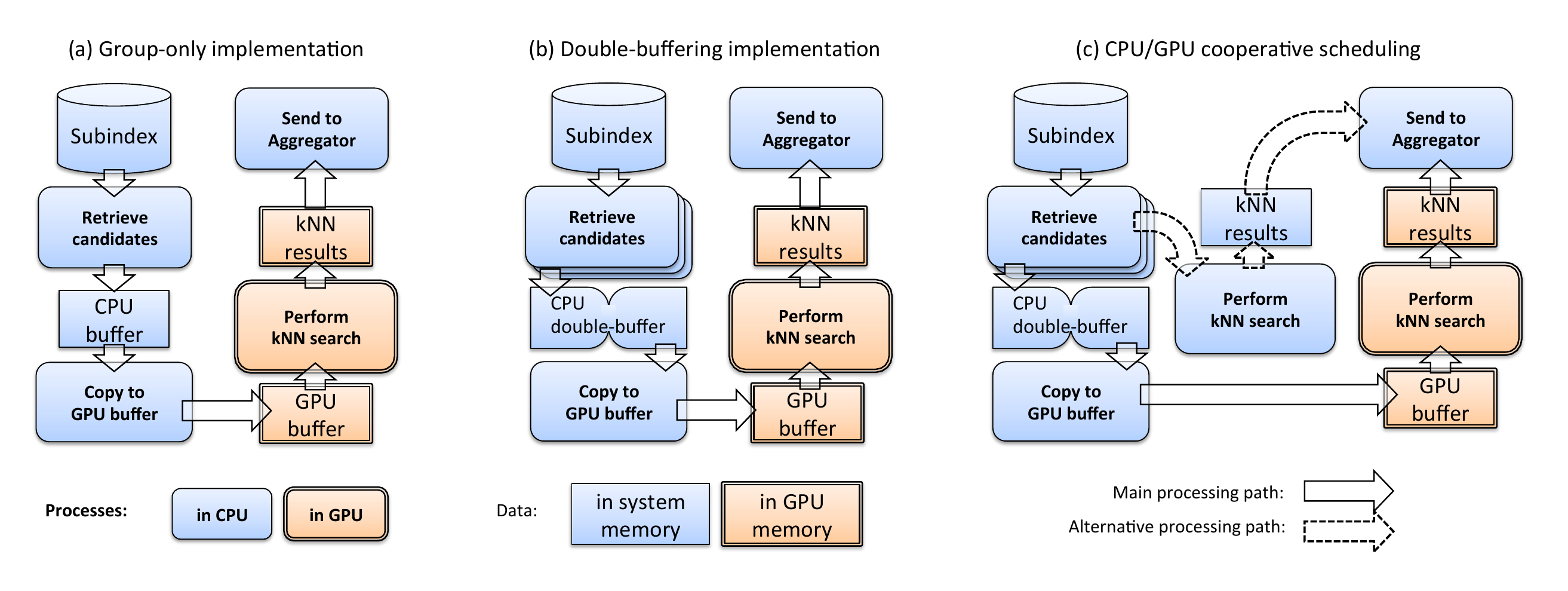}
	\caption{The three progressively sophisticated proposed schemes for implementing Hypercurves in CPU--GPU heterogeneous environments. (a)~Grouping queries to fully utilize the GPU. (b)~Additionally, employing a double-buffer to avoid idleness either on CPUs and GPU. (c)~Additionally, using CPUs unutilized capacity to perform kNN in an alternative processing path. }
\end{center}
\label{fig:parallelScheme}
\end{figure*}

In the Hypercurves pipeline, the IHLS filter is responsible for performing the
compute-intensive operations of the application. Consequently, it is the phase
of Hypercurves to be accelerated using GPUs. The computation performed by IHLS
has granularity per user request (query), and the query execution depends on
the probe-depth (the number of candidates returned from the subindexes for
further kNN computation, explained on Section~\ref{sec:multicurves}) as the
distances from the query to all candidates are calculated.

However, a single query is insufficient to fully utilize a GPU,
because probe-depths assume small values, around a few hundred
elements. Therefore, the first step towards using GPUs to accelerate this
filter was to modify IHLS to group an arbitrary number of queries
(\textbf{group-size}), which are then dispatched together for execution in
a GPU. Our parallelization uses the CPU to perform the operations related to
query grouping, while the GPU is employed during the kNN search. The execution
of the IHLS filter is divided into stages (Figure~\ref{fig:parallelScheme}--a),
explained below: 

\begin{description}

	\item[Retrieve candidates:] {returns the probe-depth vectors
closest to the query from each subindex (Lines 3 to 8 in
Algorithm~\ref{alg:searchMulticurvesDist}). Those candidates are accumulated in
a continuous block of memory (the \emph{CPU buffer}). That operation is
repeated for each query in the group. At the end, the buffer will contain
group-size sets of candidates, each set with probe-depth vectors. }

	\item [Copy to GPU buffer:] {copies the buffer from the system
memory (\emph{CPU buffer}) to the GPU memory (the \emph{GPU buffer});}

	\item [Perform kNN search:] computes the kNN search for all
group-size sets of candidates in parallel, comparing each query to its
own candidate set. At the end, returns the results in \emph{kNN results}, which
will have group-size sets, each with $k$ answers;

	\item [Send to aggregator:] {copies the results sets from the GPU
memory and sends them downstream to the Aggregator filter, which is the next
stage in processing pipeline.} 

\end{description}

The operation performed in \emph{retrieve candidates} is a binary search on
each subindex, with very irregular access patterns, dependent on both the query
and the database. Since it is not realistic to assume that an entire subindex
would fit into the GPU memory, the implementation of that stage in the
accelerator is not worthwhile: it would require intensive data
transfer between CPU and GPU. Fortunately, its computational cost is low, due to the
logarithmic growth of the binary search with respect to database size. It can be executed fairly efficiently on CPU.

\emph{Perform kNN search} is a special version of the traditional kNN, 
which compares several queries against the same database~\cite{Garcia_2008_CVGPU}. Here, however, each
query is independently compared to a different subset of the data, which
consists of the candidates just retrieved from the subindexes. In addition, the
number of queries available to execute (group-size) will vary between
executions. The group-size can be optimized according to a metric of interest,
for instance, average response time as is detailed in the
following sections. 

The kNN search itself is implemented using two GPU computing $kernels$:
(i)~\emph{CalcDist}, which calculates the distance between each query and its
candidates; (ii)~\emph{FindTopK}, which selects the $k$ nearest vectors
among the candidates, and moves them to the top positions in the list. 

The other operations, \emph{copy to GPU buffer} and \emph{send to aggregator}
involve data transfers between the system memory and the GPU memory. The latter
also involves dispatching the results for further processing downstream, in the
Aggregator filter.



\subsection{Overlapping CPU and GPU phases}
\label{sec:overlappingCPU-GPU}

In the design just discussed, the IHLS steps are performed sequentially,
resulting in no overlapping between CPU and GPU computations. However, this
approach creates idle periods in both devices. The GPU has to wait until its
data buffers are filled by the CPU with the candidates retrieved from the
subindexes; and the CPU has to wait until the kNN execution is completed on the
GPU to transfer the next batch.

In order to reduce idleness, we propose to overlap those operations by
pipelining them and using a double-buffer scheme for the communication between
operations composing the IHLS filter (Figure~\ref{fig:parallelScheme}--b). It then allows
the CPU to accumulate the sets of candidates from incoming queries, while the
GPU may be asynchronously processing the kNN search for the previous batch of
queries.  Additionally, this design employs several CPU threads to retrieve
candidates from the subindexes in parallel. That strategy significantly reduces
Hypercurves execution times, because it maximizes the utilization of the GPU.

Similar double-buffer schemes~\cite{Sancho08-doublebuffering} have being used
in multi-core CPU architectures, for instance, to overlap  useful computation
with data transfers among different levels of the memory hierarchy. Here, on
the other hand, it is employed to enable pipelining and, consequently,
to overlap execution of different code sections on different devices of the
hybrid environment.


\subsection{Cooperative execution on CPU and GPU}
\label{sec:cooperativeCPU-GPU}

In the design of Hypercurves described so far, the CPU cores perform only
operations which are not appropriate for GPUs: the retrieval of candidates, the
data transfers, and the task of coordinating the GPU execution. However, as
discussed, the cost of those CPU computations is low, and may not be sufficient to
completely occupy all CPU cores available, especially in current multi-core
architectures.
To better utilize the CPUs, we propose to employ them also to perform the kNN
computations. However that must occur only when they would otherwise be idle,
meaning that the next buffer with query candidates is ready for the GPU execution.

In our cooperative CPU/GPU scheduling solution
(Figure~\ref{fig:parallelScheme}--c), the CPU  execute both its ordinary tasks
(main processing path) and the kNN search (alternative processing path), giving
the former higher priority. Thus, the CPU will follow the first path until the
buffers used by the GPU are completely full, becoming afterwards available to
process queries on the second path. The double-buffering scheme employed
reduces the possibility of the GPU be kept waiting: even when the CPU is
momentarily held on kNN search tasks, as the CPU will usually have enough time
to fill the next buffer before the GPU is done with the current one.

The use of hybrid CPU--GPU computation, as well as the task scheduling problem
that arises of employing those processors, has received increasing attention in
the last few
years~\cite{Ding:2008:UGP:1367497.1367732,qilin09luk,mars,merge,hpdc10george,Teodoro-IPDPS2012,hartley,ravi2010compiler,Diamos:2008:HEM:1383422.1383447.Harmony}.
Those works, however, assume that implementations of all stages of the
computation are available for both devices, and try to minimize the execution
times by employing CPU--GPU tasks partition using either static
offline~\cite{mars,Ding:2008:UGP:1367497.1367732,qilin09luk} or
online~\cite{hpdc10george,ravi2010compiler} strategies. Those strategies
may work well in several contexts, including ones with divisible workloads, such as generalized reductions or MapReduce computations. However, they are restricted to cases where both CPU and GPU implementations are available for each stage. That contrasts to Hypercurves, where the CPU is used to assist the GPU in tasks that are not appropriate for acceleration, besides perform its own compute-intensive tasks during periods of idleness. Therefore, Hypercurves' task partitioning has to assign dynamically the 
priorities to types of tasks the CPU performs. 

The next sections will further elaborate on that problem of task
partition to minimize response times in online applications. To the best of
our knowledge, ours is the first work to address that problem on CPU--GPU
computation environments.

\section{Response Time Aware Task Partition in Heterogeneous CPU--GPU Platforms}
\label{sec:minQuery}

As discussed, the online nature of Hypercurves poses the interesting challenge
of optimizing the response time of the individual queries, while using
GPUs, which are throughput-oriented devices. In order to reduce response times, it is
necessary to perform a dynamic partition of the load among CPU and GPU under a
fluctuating user request load. That partition in Hypercurves is directly
related to the group-size used by IHLS, which will determine the number of tasks
queued for GPU execution and, consequently, those remaining for CPU
computation. The optimal configuration for each load intensity depends on
complex factors, including the hardware architecture, application parameters,
and dataset properties. Such complex optimization is beyond the ability of any
static configuration.

Formally, the problem can be defined as such: given a set of $n$ tasks $t_1
\dots t_n$ within a filter, $m$ processors $p_1 \dots p_m$ allocated to that
filter, the execution time of each task in each processor denoted by $e_{ij}$,
and the wait time of each task to be processed by each processor as $w_{ij}$,
we want to determine the schedule $x_{ij}$ where $1 \le i \le n$ and $1 \le j
\le m$. Each $x_{ij}$ is a binary variable indicating whether the tasks $i$ is
executed by processor $j$. Notice that, for a given task $t_i$, there exists
just one $x_{ij}$, $1 \le j \le m$ that is nonzero. Since there is no task
reordering within a processor, the $w_{ij}$ is defined as the sum of the
processing time of tasks $e_{xj}$ computed by a given processor $j$
before it gets to execute $x$. 

Our goal is to find a schedule that minimize $E$, the average execution time of
the tasks within each application filter.  For simplicity, and since the filters execute independently, we state the problem
in terms of a single filter.  In general, the execution time takes the
form:

\begin{center}
\begin{equation}
E = \operatorname{avg}_{i=0...n} \sum_{i=1}^{m} x_{ij} \times ( e_{ij} + w_{ij} )
\end{equation}
\end{center}

The execution time $e_{ij}$ varies according to the processor used. For the
GPU, it has two components: a buffer-waiting time, in which the task remains
buffered in the CPU memory; and the computation, which involves both the
data transfer between CPU and GPU, and the execution itself. The cost of both
components further depends on the task granularity (group-size).

Scheduling in such environment is difficult for various reasons: (i)~the
problem is NP-hard, since the bin packing problem, widely known to be NP-hards,
is a very simplified version of the scheduling problem discussed (the equivalence
is true when waiting times are zero); (ii)~the tasks are created at run time,
making static scheduling impossible; (iii)~estimating the execution time
($e_{ij}$) of a task has been an open challenge for decades~\cite{582071}.

The optimization is not only complex, but also dynamic, varying instantaneously as
the request load changes. Thus, exact solutions would be too costly to be
practical. We propose instead a light, greedy algorithm that calculates the
scheduling throughout the execution, optimizing the granularity of tasks
assignment to GPU (group-size), and the tasks assignment between CPU and
GPU. 

The solution we propose is driven by the idea that GPU use is advantageous only
when the aggregate throughput delivered by the available CPU computing cores is
insufficient to promptly process the demand. That approach diverges from the
usual parallelizations for heterogeneous environments, where GPUs are
systematically preferred for their capacity to achieve highest throughputs.
However, as discussed before, the use of GPUs improve throughput at the cost of
increasing the average tasks processing time. 

Therefore, it is not worthwhile using GPUs in online application unless the
tasks response times are dominated by waiting time, which occurs when the
request rates submitted to the system are above the CPU throughput. On the
other hand, under high request loads, it becomes beneficial using GPUs to
increase the system throughput: though the processing time of a single task
will be higher, due to the overheads of starting up a GPU computation, the
average response times of the system will be reduced, because the better 
throughput will lower or even eliminate waiting times.

\begin{algorithm}
\begin{small}
\caption{Dynamic Task Assigner for Heterogeneous Environments --- DTAHE.}
\label{alg:dtaa}
\flushleft
Executes in parallel with the other threads, until the filter has nothing else to process ($\mathit{EndOfWorks}$ becomes true). There is one independent instance of this thread for each transparent copy of each filter.
\begin{description}
	\item[$\mathit{EndOfWorks}$:]        true if this filter copy has no more processing to do, false otherwise
	\item[$\mathit{CC}$:]                number of CPU cores allocated to this filter copy
	\item[$\mathit{NumEvents}$:]         the number of events waiting to be processed
	\item[$\mathit{GetNextEvent}()$:]    pops from the waiting queues the next event to be processed
	\item[$\mathit{CurrentBuffer}$:]     current GPU buffer, the one being prepared
	\item[$\mathit{ProcessInCPU}(e)$:]   dispatches the event $e$ to be processed in the CPU
	\item[$\mathit{BufferEvent}(e, b)$:] stores the event $e$ to be processed in the buffer $b$	
	\item[$\mathit{IsGPUIdle}$:]         true if the GPU is idle, false otherwise
	\item[$\mathit{IsFull}(b)$:]         returns true if the buffer $b$ specified is full, false otherwise
	\item[$\mathit{QueueInGPU}(b)$:]     queue the buffer $b$ to be processed in the GPU
\end{description}
\begin{algorithmic}[1]
\WHILE {\NOT $\mathit{EndOfWork}$}
	\STATE {$\mathit{ready} \leftarrow \mathit{NumEvents}$}
	\IF { $\mathit{ready} \leq \mathit{CC}$ \textbf{or} $\mathit{CurrentBuffer}$~\textbf{is full}}
			\STATE {$\mathit{ProcessInCPU}(\mathit{GetNextEvent}())$ }
	\ELSE
		\STATE {$\mathit{event} \leftarrow \mathit{GetNextEvent}()$}
		\STATE {$\mathit{BufferEvent}(\mathit{event}, \mathit{CurrentBuffer})$}
		\STATE {$\mathit{ready} \leftarrow \mathit{NumEvents}()$}
		\IF { ($\mathit{ready} < \mathit{CC}$ \textbf{and} $\mathit{IsGPUIdle}$) \textbf{or} $\mathit{IsFull}(\mathit{CurrentBuffer})$}
			\STATE $\mathit{QueueInGPU}(\mathit{CurrentBuffer})$
		\ENDIF
	\ENDIF
\ENDWHILE
\end{algorithmic}
\end{small}
\end{algorithm}

The solution we propose is called Dynamic Task Assigner for Heterogeneous
Environments (DTAHE) and is presented in Algorithm~\ref{alg:dtaa}. The DTAHE is
executed in parallel with the computing threads in each filter copy, and loops until
the upstreams filters notify that execution has finished --- the
$\mathit{EndOfWorks}$ message. Line 2 of the algorithm will check the number of
events/requests queued in the filter ($\mathit{ready}$), and further decide
whether the thread should follow the event processing using the CPU or the GPU (line 3). 

If the GPU buffer is full or if the CPU computing cores available are enough to
process all queued events, an event is dequeued and dispatched to the CPU (line
4). Otherwise, an event is dequeued, the items to be compared to that query
are retrieved from the subindexes and stored in the GPU data buffer (Lines 6--7).
The scheduler decides, then, if the buffer should be sent for the execution in
the GPU, which happens when it becomes full, when the GPU becomes idle (which
we want to avoid), or if the number of ready events becomes low enough so that
the remaining ones can be processed in the CPUs. Each GPU has one CPU computing
core reserved for managing purposes (coordination, buffer copying, etc.), but,
when that GPU has no buffers ready to be processed, its dedicated CPU core
become available to perform other tasks. The motivation, again, is to get the
GPU running as often and as early as possible, avoiding as much as possible to
keep it waiting, while keeping the CPU available to process some queries when
the GPU is busy.

DTAHE solves at once the problem of dynamically distributing the tasks among
CPUs and GPUs and determining the optimal group-size. The latter is done
implicitly, as buffers are dispatched for GPU computation either when they are
full (maximum group-size, corresponding to situations of high loads and maximum
throughput) either when the GPU becomes idle (smaller group-sizes, as the load
becomes lighter). When the load becomes light enough to be dealt with just the
CPU cores, the GPU is not employed, again keeping response times as short as
possible. On the other hand, as the load increases, processing migrates
increasingly to the GPU.

The problem of optimizing group-sizes is similar to optimizing the parallelism
granularity in parallel loops or \emph{doall}-like operations, which have been
extensively studied~\cite{Chen:1990:ISG:325096.325150}. Most works on that
area, however, focus on applying loop transformations according to the
available resources, in order to adjust the granularity, reducing the
synchronization overhead and do not take into account load variability. The
best parallelism  is often beyond any static tuning. That has motived several
recent works, which focus on runtime
transformations~\cite{Aleen:2010:IDE:1693453.1693494,Blagojevic:2007:RSD:1316088.1316319,springerlink:10.1007/978-3-540-31832-3-12}.
Those interesting works aim at parallelism tuning for a lower level and are
only concerned with CPU-based machines, thus being complementary to the strategy we
propose.

\section{Experimental results}
\label{sec:results}

In this section, we evaluate the impact of our propositions on Hypercurves'
performance. The experiments have been performed using two setups of machines.
The first setup consisted of 2 PCs connected through a Gigabit Ethernet, each
with two quad-core AMD Opteron 2.0~GHz 2350, 16~GB of main memory, and one
NVidia GeForce GTX260 GPU. The second setup was a 8-node cluster connected with
Gigabit Ethernet, each node being a PC with two quad-core hyper-threaded Intel
Xeon E5520, 24~GB of main memory, and one NVidia GeForce GTX470. All machines
run Linux. 

The main database used to evaluate our algorithm contained 130,463,526 SIFT
local feature vectors~\cite{Lowe:2004:DIF:993451.996342}, with 128 dimensions
each. Those feature vectors have been computed from 233,852 background images
from the Web, and 225 foreground images from our personal collections. The
foreground images have been used to compute sets of feature vectors that must
be matched, while the background images have generated the feature vectors used
to confound the method. The foreground images, after strong transformations
(rotations, changes in scale, non-linear photometric transformations, noise,
etc.) have also been used to create 187,839 query feature vectors.  Due to the
number of evaluations performed, we have also employed smaller partitions of
that main database, in order both to achieve feasible experimentation running
times, and to emphasize certain aspects (e.g., overhead) in specific
experiments.  

The experiments concentrate on issues of efficiency, since, as demonstrated in
Section~\ref{sec:equivalence}, Hypercurves, with very high probability, has the
same results of Multicurves. Thus, by construction, it inherits the good
trade-off between precision and speed of
Multicurves~\cite{Valle:2010:10.1109:TCE.2010.5606242}. 

\subsection{The Impact of Task Granularity}
\label{sec:task-impact}

As discussed in Section~\ref{sec:gpuIHLS}, task granularity has an important
impact on the GPU acceleration achieved. In Hypercurves, it is
dictated by the group-size: the number of queries aggregated to be processed
concurrently within each GPU execution (Section~\ref{sec:gpuIHLS}). 

This study has employed a subsample of 1,000,000 feature vectors randomly
selected from the main database, and 30,000 queries. A smaller dataset has been
used in order to provide shorter execution times, which are more appropriate to
highlight any potential overheads in our solution. The entire set of queries
has been dispatched to the filter QR at the beginning of the execution,
creating a high number of concurrent queries ready to execute in the IHLS
filter throughout execution (check the filter enchainment in
Figure~\ref{fig:pmulticurves}). In addition, in this initial evaluation, a
single machine has been employed.

\begin{figure*}[htb!]
\begin{center}
\mbox{
\subfigure[2-node AMD --- Probe-depth=250]{\label{fig:opt250amd}
        \includegraphics[width=0.44\textwidth]{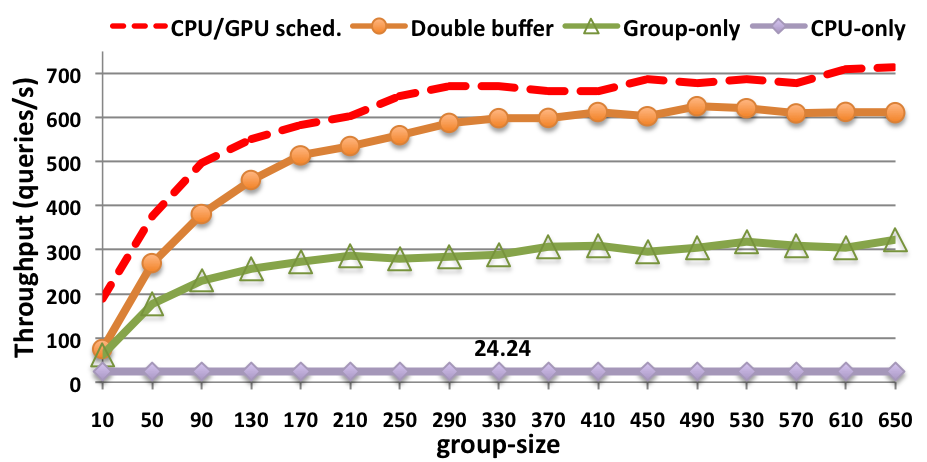}}
}
\mbox{
\subfigure[8-node Intel --- Probe-depth=250]{\label{fig:opt250intel}
        \includegraphics[width=0.44\textwidth]{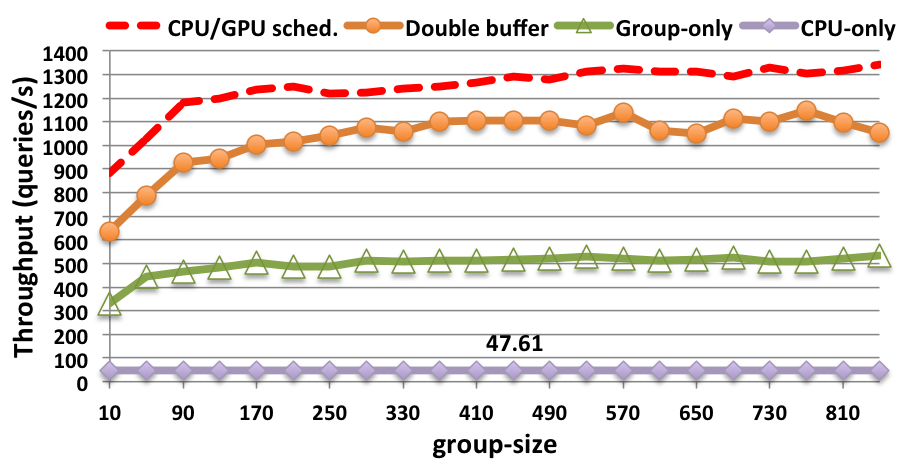}}
}          
\mbox{
\subfigure[2-node AMD --- Probe-depth=350]{\label{fig:opt350amd}
        \includegraphics[width=0.44\textwidth]{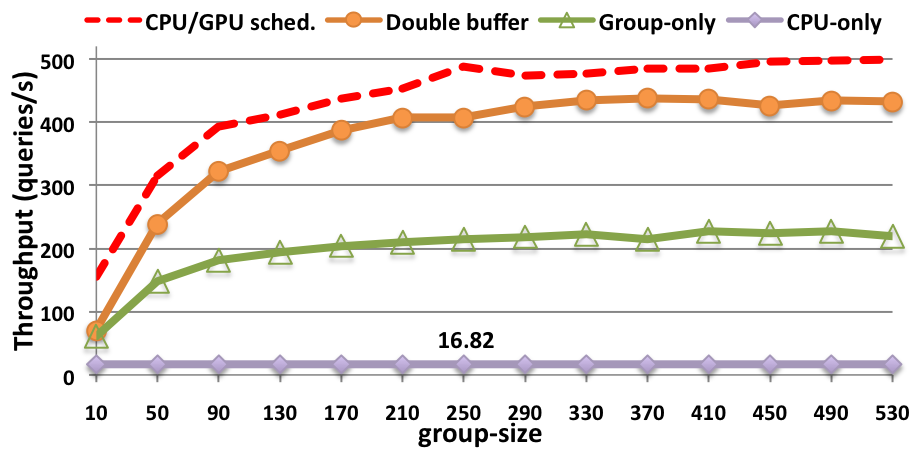}}
}
\mbox{
\subfigure[8-node Intel --- Probe-depth=350]{\label{fig:opt350intel}
        \includegraphics[width=0.44\textwidth]{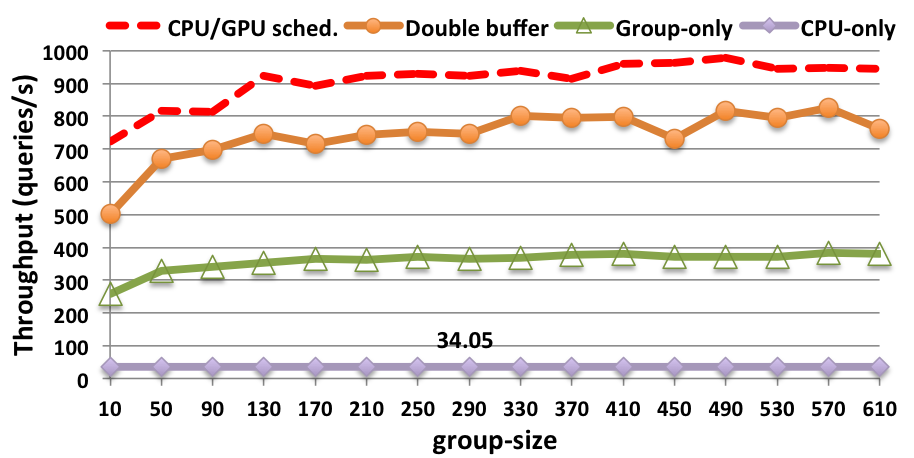}}
}

\mbox{
\subfigure[2-node AMD --- Probe-depth=450]{\label{fig:opt450amd}
        \includegraphics[width=0.44\textwidth]{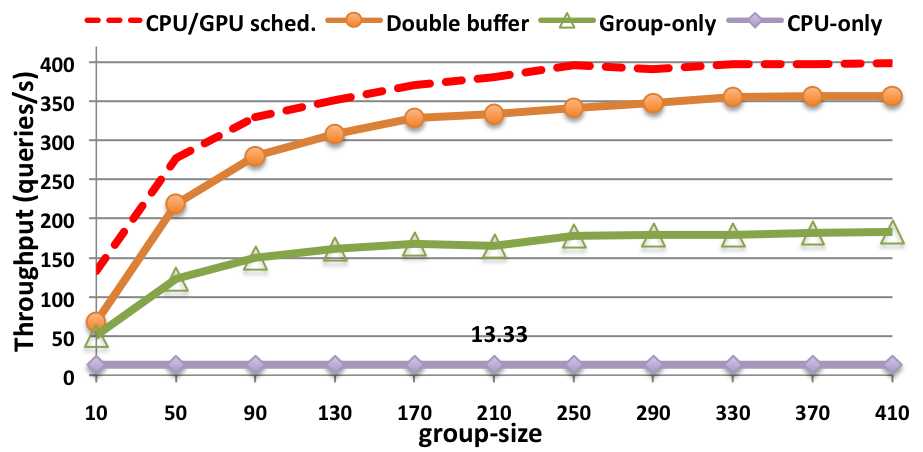}}
} 
\mbox{
\subfigure[8-node Intel --- Probe-depth=450]{\label{fig:opt450intel}
        \includegraphics[width=0.44\textwidth]{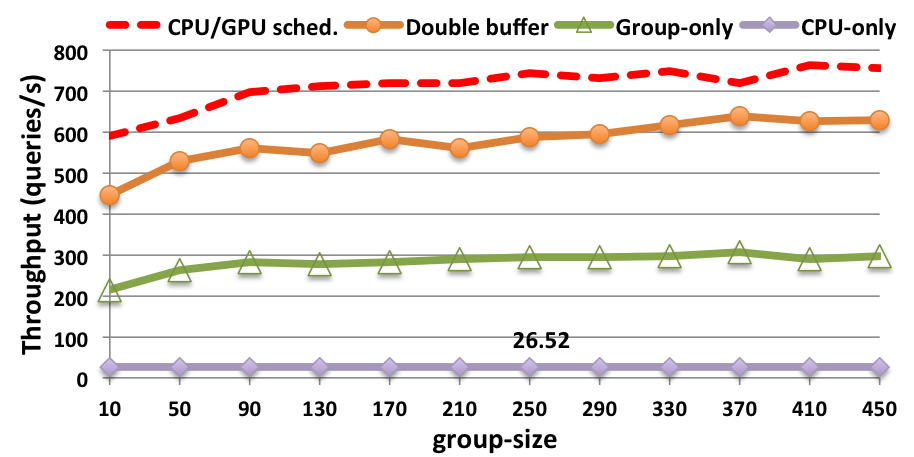}}
}          
\caption{Hypercurves performance as group-size varies, for multiple probe-depth values, in both machine configurations (2 nodes of AMD/GTX260, and 8 nodes of Intel/GTX450). Dynamic scheduling is \emph{not} employed in those experiments: the parameters are kept fixed on each execution (each data point). The speedup brought by the use of the GPU is dramatic, and each successive improvement (from group-only to double-buffering to cooperative CPU/GPU scheduling) is considerable.}
\label{fig:optsSpeedup}
\end{center}
\end{figure*}

Figure~\ref{fig:optsSpeedup} presents Hypercurves throughput (in queries per
second), for 3 typical choices of probe-depth, as group-size varies up to the
limit of queries that can be accommodated within the GPU memory. Group-size is
kept fixed during each single execution.  We focus on the impact of task
grouping, expressed on the ``Grouping only'' curves in the graphs. The other
results presented in the same graphs are discussed on the following sections.

As expected, a small number of queries is insufficient to completely utilize
the GPU, but the GPU performance is consistently improved as group-size
increases. Also, when 50\% of the maximal group-size is employed, the speedups
achieved are about 90\% of the best acceleration.  In addition, the best
speedups are similar for all probe-depths: 13.34$\times$ (probe-depth=250),
13.52$\times$ (probe-depth=350), and 13.73$\times$ (probe-depth=450). The small
impact of the probe-depth over the speedup is an important, positive property of
the method, since it allows the user to calibrate that parameter more or less
freely, provided that there are enough queries to group and process in
parallel. 

The gains attained by task grouping are promising, enhancing
a very efficient approximate search algorithm --- whose sequential CPU
implementation is already orders of magnitude faster than the exact search. In
the following sections, we will demonstrate increasing performance gains, obtained
by the optimizations proposed on top of that GPU parallelization (See
Section~\ref{sec:heterogeneous}).

\subsection{The Effect of Overlapping CPU and GPU}

Figure~\ref{fig:optsSpeedup} presents the performance of Hypercurves when using
the double-buffering scheme intended to maximize the GPU utilization
by reducing the waiting time between batches of queries. That enhancement, discussed in
Section~\ref{sec:overlappingCPU-GPU}, is built on top of the grouping mechanism demonstrated in the
previous section. The results are shown in the curve labeled ``Double buffer''.

Similarly to the ``Grouping only'' case, the performance grows as the value of group-size
increases, nearly doubling the throughput of the grouping-only version for most
of the group-size values. When compared to the single CPU core execution, the
maximum speedups of the double-buffered GPU-accelerated version are 25.81,
26.02, and 26.75, for probe-depths 250, 350, and 450 respectively.

During the evaluation, a complete overlapping of the CPU and the GPU was
achieved, except at the very beginning, during the preparation of the first
batch of queries. We have also observed that, in all experiments, the GPU was
always occupied processing queries, while the CPU experienced idle periods for
the threads responsible for retrieving candidates from the subindexes and copy them
to the buffers. The CPU underutilization motivated its use in the kNN phase of the application, 
in addition to the preparation of the buffers for the GPU. That cooperative strategy is evaluated in the
next section.

\subsection{Maximizing CPU--GPU Cooperation}
\label{sec:maxcpugpu}

Since the CPUs present in our systems prepare the buffers faster than the GPU
is able to consume them, they experience idle periods. To maximize the system
performance under those circumstances, we also employed the CPU cores in the
compute-intensive kNN tasks during that idleness, as discussed in
Section~\ref{sec:cooperativeCPU-GPU}. In the remainder of this section, the
benefits of that strategy are evaluated in addition to double-buffering and
grouping. Figure~\ref{fig:optsSpeedup} shows the results, in the curves
labeled ``CPU/GPU sched.''. 

As shown, the gains with that technique are very relevant, and
speedups of 1.23$\times$, 1.22$\times$ and 1.22$\times$ for, respectively,
probe-depths of 250, 350 and 450 were achieved on top of the previous version
of Hypercurves.

\begin{figure}[htb!]
\begin{center}
        \includegraphics[width=0.49\textwidth]{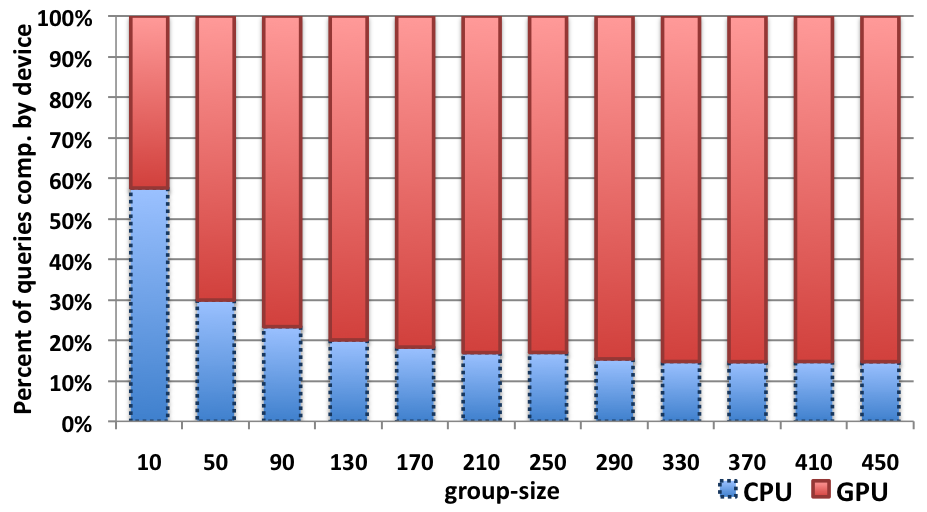}
\caption{Fraction of the tasks computed by each type of device (CPU vs. GPU) as group-size varies, illustrating how GPU utilization is favored by large group-sizes. Those experiments were executed on an AMD node using probe-depth of 350.}
\label{fig:cpugpuload}
\end{center}
\end{figure}

Interestingly, the speedup obtained here slightly decreases as the group-size increases. This behavior is
consequence of the higher efficiency of the GPU for large group-sizes, which
tends to reduce the idle time of the CPU cores. Figure~\ref{fig:cpugpuload}
illustrates that as the group-size grows, more queries tend to be processed by
the GPU instead of the using CPU for the kNN phase.  Even so, the improvements achieved were 
rewarding, with more than 1.2$\times$ average speedup across the group-size configurations used. As the number of available
CPU cores also tend to increase in new architectures, the potential of using
those devices in cooperation with the GPU cannot be neglected.

\subsection{Granularity vs. Response Times}
\label{sec:granu}

So far, in the experimental evaluation, we have analyzed Hypercurves performance
in scenarios where a very large number of queries is submitted at once,
thus assessing the application throughput capabilities. However, in real-world 
operation, the query rate submitted to an online application is generated by users, and is subjected to variability throughout the
execution. Moreover, under those circumstances, the most important metric is
the response time observed by the users. Therefore, in this section, we analyze
how the grouping technique and cooperative CPU--GPU execution impact the
average query response time.

\begin{figure}[htb!]
\begin{center}
\mbox{
\subfigure[2-node AMD --- GPU-only]{
\label{fig:queryratefixed-octopus}
        \includegraphics[width=0.49\textwidth]{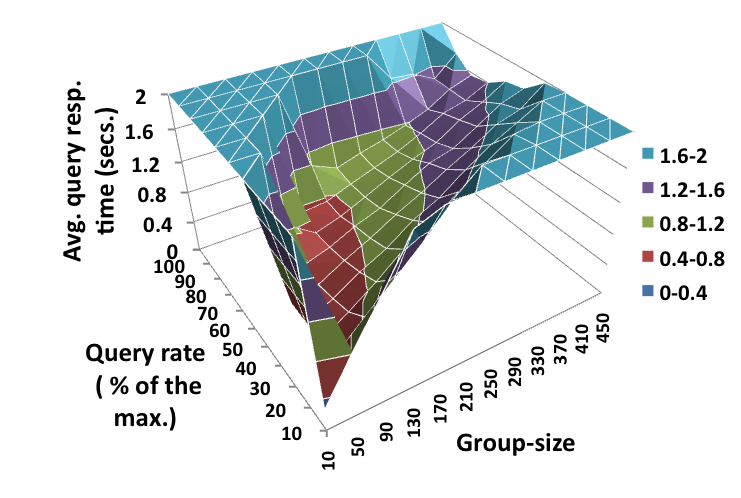}
}
}
\mbox{
\subfigure[2-node AMD --- CPU--GPU]{\label{fig:queryratefixed-octopus-cpu-gpu}
        \includegraphics[width=0.49\textwidth]{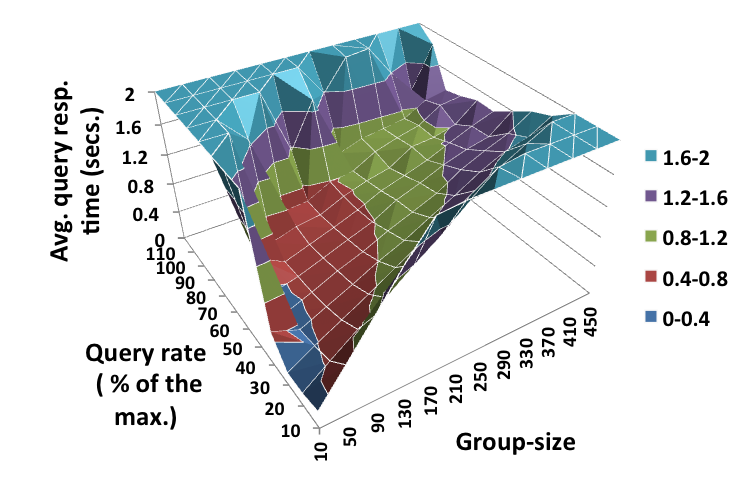}}
}
\mbox{
\subfigure[8-node Intel --- CPU--GPU]{\label{fig:queryratefixed-hydra}
        \includegraphics[width=0.49\textwidth]{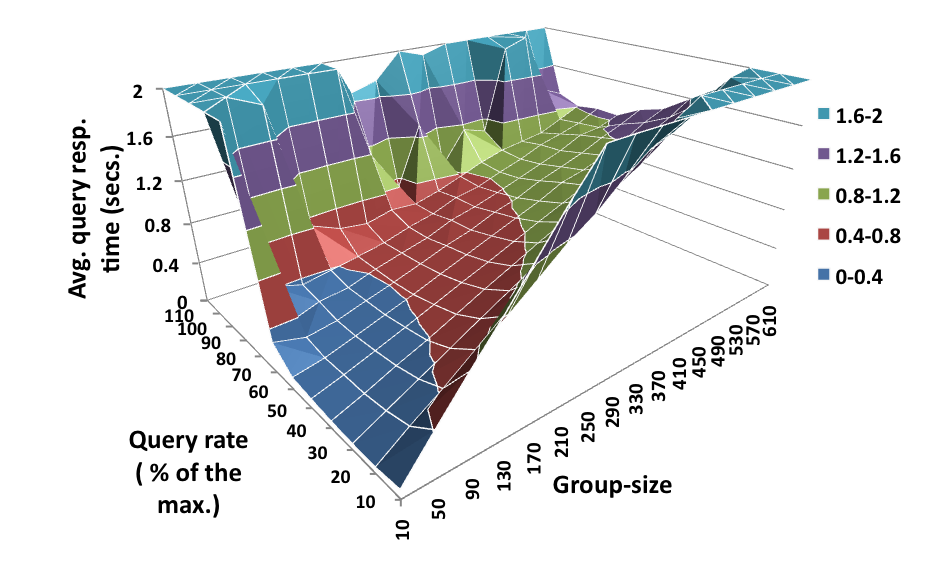}}
}       
\vspace{-3mm}
\caption{Average query response times as query rate (\% of the maximum) and group-size vary, for a probe-depth of 350. The parameters are kept fixed on each execution (each data point). The graphs demonstrate the complex task of optimizing group-size as the load (query-rate) fluctuates.}
\label{fig:queryratefixed}
\end{center}
\vspace{-3mm}
\end{figure}

In this analysis, we vary both the number of queries submitted per second and
group-size across experiments, but \emph{keep them fixed within each run}. The
capacity of adjusting the group-size dynamically is evaluated in the next
section.

We employ the same 1,000,000-vectors database, and the 30,000 queries used in
previous sections. First, in Figure~\ref{fig:queryratefixed-octopus}, we
present the average response times of Hypercurves when the GPU is used to
compute all the tasks, and the group-size is varied. 

It is noticeable that a single, fixed, group-size value is unable to
deliver the best response times for all request loads. The response-time
function we seek to optimize has a complex behavior and its minimum moves up
the group-size axis as the query rate increases.  Hypercurves has been able to
answer queries in reasonably low response times until the load reached about
80\% of the maximum supported by the configuration, but after that point
response times have grown steeply.

The comparison of Hypercurves using only the GPU for kNN, versus the
cooperative CPU--GPU configuration is possible contrasting
Figures~\ref{fig:queryratefixed-octopus}
and~\ref{fig:queryratefixed-octopus-cpu-gpu}. Visual inspection of those
figures show that the CPU--GPU cooperation resulted in a systematic reduction
of the average response times.  Moreover, across all the experiments, the
CPU--GPU version reduces the response times in 58\%, on average. Those gains are much
more impacting that the previous improvements in throughput achieved with the
same cooperation (Section~\ref{sec:maxcpugpu}). The reason is that although the
CPU has a lower throughput, the response times of the queries processed by this
device are much smaller: the queries routed to the GPU have a longer processing
path, having to be queued, copied to the GPU, processed using lower frequency
cores, and copied back to the system main memory.

We also present the performance of Hypercurves using CPU--GPU collaboratively
for the Intel/GTX470 node in Figure~\ref{fig:queryratefixed-hydra}.  When
compared to the AMD/GTX260 machine, the performance of the Intel node is superior, and
much better average response-times are achieved. That is mainly a consequence
of the improvements in design achieved by the GTX470 GPU as compared to the
GTX260 GPU, as the first has more computing cores, better bandwidth, etc.

\subsection{Response times on variable request rates}
\label{sec:variable}

In this section, we analyze the DTAHE dynamic scheduler
(Section~\ref{sec:minQuery}) capacity to adapt Hypercurves' work partition
among CPU and GPU under scenarios with stochastically variable workloads. During
this evaluation, the load/request rate follows a Poisson distribution, with
expected average rate ($\lambda$) varying from 20 to 100\% of the maximum
throughput of the application in each configuration. That maximum throughput is
computed in a preliminary run, where all the queries are sent for computation in
the beginning of the execution. This set of experiments employ the cooperative CPU--GPU computation, and a
probe-depth of 350.

\begin{table*}[htb!]
\begin{center}
\subtable[First setup (2-node AMD)]{\label{fig:amd-query-resp}
\begin{tabular}{c|c|c|c|c|c}
\hline
\multirow{2}{*}{\textbf{Scheduling}} & \multicolumn{5}{|c}{\textbf{ Poisson $\lambda$ (\% of max. throughput)}} \\ \cline{2-6}
 			& 20	& 40 	& 60 	 & 80 	 & 100 	 	\\ \hline
Best static	& 0.11	& 0.4 	& 0.42  & 0.61 & 0.98		\\ 
DTAHE		& 0.06	& 0.13	& 0.14  & 0.22 & 1.02		\\ \hline
\end{tabular}
}
\subtable[Second setup (8-node Intel)]{\label{fig:hydra-query-resp}
\begin{tabular}{c|c|c|c|c|c}
\hline
\multirow{2}{*}{\textbf{Scheduling}} & \multicolumn{5}{|c}{\textbf{ Poisson $\lambda$ (\% of max. throughput)}} \\ \cline{2-6}
			& 20	& 40 	& 60 	 & 80 	 & 100 	 	\\ \hline
Best static & 0.054	& 0.098 & 0.12   & 0.25 & 0.65	\\ 
DTAHE		& 0.034	& 0.089	& 0.1    & 0.16 & 0.68		\\ \hline

\end{tabular}
}
\end{center}
\caption{Average query response times (in s) for static and dynamically-tuned scheduling configurations, under stochastic loads. Unless the system is completely saturated, the dynamic scheduling always wins, usually by a considerable margin.}
\label{tab:query-resp}
\end{table*}

The results are summarized in Table~\ref{tab:query-resp}.  The best static
configuration refers to the minimum response times achieved from an exhaustive
search in the group-size space for each query rate employed. It is
noticeable that DTAHE strongly outperform the best static configuration for
most of the cases. On average, its response times are 52\% (2-node AMD) and 81\% (8-node Intel) of the best
static configuration. 

The only configurations where DTAHE falls slightly behind static scheduling have Poisson rates
$\lambda$ equal to 100\% of the maximum throughput bearable by the machines. In
that extreme scenario not much can be accomplished by dynamic scheduling, since
the goal of maximum throughput coincides with the one of minimum response
times, and a simple static configuration manages that with slightly less
overhead. The differences, however, are small, and
dynamic scheduling is, of course, much more flexible.

\subsection{Evaluating Hypercurves' scalability}
\label{sec:scale}

The distributed memory analysis in this section has focused on evaluating
Hypercurves scaleup. We consider the compromises between performance of the
parallelism and conservation of the results precision, as the database is
partitioned among the computing nodes. The scaleup evaluation is appropriate in
our application scenario because we expect to have an abundant volume of data for
indexing, thus the speedup evaluation starting with a single node holding the
entire database might not be realistic. The experiments executed in this
section used the 8-node Intel cluster, employing all CPU cores and GPUs
available on the nodes. The main database with 130,463,526 local feature
vectors is used proportionally, with $n/8$ of the database being used for the
experiment with $n$ nodes.

\begin{table*}[htp!]
\begin{center}
\begin{small}
\begin{tabular}{c|c|c|c|c|c|c|c|c}\hline
{\bf Number of nodes} 			& 1  	& 2  	& 3  	& 4  	& 5  	& 6  	& 7   	& 8 	\\ 
{\bf \# of cores / \# of GPUs} 		& 16 / 1 	& 32 / 2 	& 48 / 3 	& 64 / 4 	& 70 / 5 	& 86 / 6 	& 102 / 7 & 118 / 8 \\ \hline
{\bf Optimist --- queries per s}   	& 964   & 1904  & 2649 	& 3598 	& 4490 	& 5397 	& 6297	& 7483	\\ {\bf Pessimist --- queries per s}   	& 964  	& 1683  & 2197  & 2849  & 3416 	& 3968 	& 4498	& 5135	\\ \hline
\end{tabular}
\caption{Scaleup evaluation: query rate as database and number of nodes increase proportionally (probe-depth=350).}
\label{tab:scaleup}
\end{small}
\end{center}
\end{table*}

The query rate delivered by the algorithm considers two parameterization
scenarios named Optimist and Pessimist (Table~\ref{tab:scaleup}), which differ
in their guarantees of equivalence (in terms of precision of the kNN search) to
the sequential Multicurves algorithm. The Optimist parameterization divides the
probe-depth equally among the nodes, without any slack --- it will only be
equivalent to Multicurves in the unlikely case that all candidates of that
query are equally distributed on the nodes.  The Pessimist parameterization
uses a slack that guarantees a probability smaller than 2\% that a candidate
vector selected by the sequential algorithm will be missing from the distributed
version (see Section~\ref{sec:equivalence} for details). Note that that choice
is extremely conservative, because in order to effectively affect the answer,
the missed feature vectors from the candidate set have to be among the actual top-$k$
set, and $k$ is much smaller than the probe-depth. 

Table~\ref{tab:scaleup} presents Hypercurves query rates on the scaleup
evaluation. As shown, the scalability of the algorithm is impressive for both
Optimist and Pessimist configurations, achieving \emph{superlinear scaleup} in
all setups. That strong performance of Hypercurves is observed because the
application is only affected by the size of the database during the phase where
candidates are retrieved from subindexes and the cost of that stage grows only
logarithmically with the size of the database. The costly phase of computing
the distances from the query to the retrieved candidates can, thanks to the
probabilistic equivalence (Section~\ref{sec:equivalence}), be efficiently
distributed among the nodes, with a relatively small overhead.

Not only is the scalability of the algorithm very good, but its raw response
rates (queries per s) are very high. For instance, number of queries that the
algorithm would be able to answer a per day are: 646 and 443 million,
respectively, for the Optimist and the Pessimist configurations. Those rates
indicate that by employing the technology proposed, a large-scale image search
system could be built at reasonably low hardware and power costs per request, 
since GPU accelerators are very computational- and power-efficient platforms.

\section{Conclusions and future work}
\label{sec:conclusions}

This work has proposed and evaluated Hypercurves, an online similarity search
engine for very large multimedia databases. 
Hypercurves has been designed to fully exploit massively parallel  machines, with
both CPUs and GPUs. Its use in a CPU--GPU environment, along with
a set of optimizations, resulted in accelerations of about 30$\times$ on top of
the single-core CPU version. 

We have also studied the problem of response-time aware (DTAHE)
partition of tasks between CPU and GPU, under request load fluctuations, which
occurs as a result of the variating number of queries submitted by the user to
the application. DTAHE has been able to reduce the average query response times
in about 50\% and 80\% (respectively for both machine configurations used in the experiments), when compared to the
best static partition in each case. Furthermore, Hypercurves achieved
\emph{superlinear} scaleups in all experiments, while keeping a high guarantee
of equivalence with the sequential Multicurves algorithm, as asserted by the
proof of probabilistic equivalence.

We are currently interested in the complex interactions between algorithmic
design and parallel implementation for services such as Hypercurves. We are
also investigating how a complete system for content-based image retrieval can
be built upon our indexing services, and optimized using our techniques and
scheduling algorithms. We consider it a promising direction for future, as
Hypercurves subindexes implementation in heterogeneous environments offers very
good reply rate.

\balance
\bibliographystyle{spmpsci}      
\bibliography{george,eduardo}   

\end{document}